\documentclass[10pt]{iopart}
\usepackage{graphicx}
\usepackage{bm}
\eqnobysec

\newcommand{\beq}{\begin{equation}}
\newcommand{\beqa}{\begin{eqnarray}}
\newcommand{\eeq}{\end{equation}}
\newcommand{\eeqa}{\end{eqnarray}}
\newcommand{\bra}[1]{\langle#1\vert}
\newcommand{\ket}[1]{\vert#1\rangle}
\newcommand{\braket}[2]{\langle#1\vert#2\rangle}
\newcommand{\braopket}[3]{\langle#1\vert#2\vert#3\rangle}
\newcommand{\abs}[1]{\vert#1\vert}
\newcommand{\bdy}{_{\rm B}}
\newcommand{\cotanh}{{\mathop{\rm cotanh}\nolimits}\,}
\newcommand{\dd}{{\rm d}}
\newcommand{\eps}{\varepsilon}
\newcommand{\erf}{{\mathop{\rm erf}\nolimits}\,}
\newcommand{\euler}{{\gamma_{\scriptstyle{\rm E}}}}
\newcommand{\frad}[2]{{\displaystyle{\displaystyle#1\over\displaystyle#2}}}
\newcommand{\half}{{\scriptstyle{\scriptstyle1\over\scriptstyle2}}}
\newcommand{\ii}{{\rm i}}
\newcommand{\imax}{_{\rm max}}
\newcommand{\imin}{_{\rm min}}
\newcommand{\m}[1]{{\bm#1}}
\newcommand{\mean}[1]{\langle#1\rangle}
\newcommand{\new}{_{\rm new}}
\newcommand{\odd}{\,{\rm odd}}
\newcommand{\typ}{_{\rm typ}}
\newcommand{\un}{^{(1)}}
\newcommand{\w}[1]{{\widetilde{#1}}}
\newcommand{\ze}{^{(0)}}
\newcommand{\A}{{\cal A}}
\renewcommand{\H}{{\cal H}}
\renewcommand{\Im}{{\mathop{\rm Im}\nolimits}\,}
\renewcommand{\Re}{{\mathop{\rm Re}\nolimits}\,}
\newcommand{\T}{{\cal T}}
\newcommand{\Tbar}{{\overline T}}

\begin{document}

\title[Nonequilibrium stationary states of Ornstein-Uhlenbeck processes]
{Characterising the nonequilibrium stationary states of Ornstein-Uhlenbeck processes}

\author{Claude Godr\`eche and Jean-Marc Luck}

\address{Institut de Physique Th\'eorique, Universit\'e Paris-Saclay, CEA and CNRS,
91191~Gif-sur-Yvette, France}

\begin{abstract}
We characterise the nonequilibrium stationary state
of a generic multivariate Ornstein-Uhlenbeck process involving $N$ degrees of freedom.
The irreversibility of the process is encoded
in the antisymmetric part of the Onsager matrix.
The linearity of the Langevin equations allows us to derive closed-form expressions
in terms of the latter matrix for many quantities of interest,
in particular the entropy production rate and the fluctuation-dissipation ratio matrix.
This general setting is then illustrated by two classes of systems.
First, we consider the one-dimensional ferromagnetic Gaussian spin model
endowed with a stochastic dynamics where spatial asymmetry results in irreversibility.
The stationary state on a ring is independent of the asymmetry parameter,
whereas it depends continuously on the latter on an open chain.
Much attention is also paid to finite-size effects, especially near the critical point.
Second, we consider arrays of resistively coupled electrical circuits.
The entropy production rate is evaluated in the regime
where the local temperatures of the resistors have small fluctuations.
For $RL$ networks
the entropy production rate grows linearly with the size of the array.
For $RC$ networks
a quadratic growth law violating extensivity is predicted.
\end{abstract}

\ead{\mailto{claude.godreche@ipht.fr,jean-marc.luck@ipht.fr}}

\maketitle

\section{Introduction}
\label{intro}

The everlasting interest for nonequilibrium phenomena in Statistical Physics
has experienced a considerable upsurge in the last decades
with the emergence of a series of general results known as fluctuation theorems.
These theoretical developments have emphasised the central role
of the rate of entropy production per unit time,
which is the key quantity involved
in the Gallavotti-Cohen fluctuation theorem~\cite{GC}
(see also~\cite{ELS,K,CR,LS,maes,SE}).
Even though the entropy production rate appears as a fundamental quantity
characterising nonequilibrium stationary states,
the latter possess many other features of interest,
including a violation of the fluctuation-dissipation theorem
characteristic of equilibrium states in the linear-response regime
and non-local fluctuations
in the large-deviation regime~\cite{Derrida,MPR,Touchette,Seifert}.

The goal of this work is to investigate the most salient characteristic features
of nonequilibrium stationary states within a unified framework
in a specific class of models which lend themselves to an analytic approach,
namely linear Langevin systems.
These systems can be viewed as higher-dimensional extensions
of the well-known Ornstein-Uhlenbeck process~\cite{OU},
describing the velocity of a Brownian particle in one dimension,
\beq
m\frac{\dd v(t)}{\dd t}=-\gamma v(t)+\eta(t),
\eeq
where $\gamma$ is the friction coefficient
and $\eta(t)$ is a Gaussian white noise with amplitude~$\sigma$, i.e.,
\beq
\mean{\eta(t)\eta(t')}=2\sigma^2\delta(t-t').
\eeq
The system equilibrates under the combined effects of linear damping and noise.
The equilibrium probability distribution of the velocity
is the Maxwell-Boltzmann distribution, i.e., a Gaussian law with variance
$\mean{v^2}=\sigma^2/(m\gamma)=kT/m$.
The Ornstein-Uhlenbeck process is reversible
and obeys detailed balance with respect to the above distribution.
The Einstein relation $\sigma^2=kT\gamma$ is an example
of an equilibrium fluctuation-dissipation relation.

Generalising the Ornstein-Uhlenbeck process to several linearly coupled degrees of freedom
appears as a very natural construction.
The first occurrence of a multivariate Ornstein-Uhlenbeck process
can be traced back to~\cite{WU}.
Such a process is defined by~$N$ linear Langevin equations
describing the evolution of coupled dynamical variables.
Remarkably enough, in stark contrast with the historical case recalled above,
a generic multivariate Ornstein-Uhlenbeck process is irreversible~\cite{L,lax66,R,VK,G}.
To close this introduction of the problem,
let us mention related works concerning
the structure of either dissipative or otherwise irreversible dynamical systems
near fixed points, where the dynamical equations can be linearised~\cite{GT,AO,KAT,HQ2}.
In the stochastic setting,
this line of thought leads to consider multivariate Ornstein-Uhlenbeck processes.

This paper is devoted to a systematic characterisation
of the nonequilibrium stationary states of multivariate Ornstein-Uhlenbeck processes.
The linearity of the Langevin equations governing these processes
allows us to use tools from linear algebra.
In section~\ref{general}, devoted to the general framework,
we demonstrate that the amount of irreversibility
is measured by the antisymmetric part $\m Q$ of the Onsager matrix~$\m L$.
Multivariate Ornstein-Uhlenbeck processes are defined in section~\ref{model},
the Sylvester equation governing their steady state is derived in section~\ref{steady},
and their irreversibility is characterised in section~\ref{irreversible}.
The various quantities characterising
their nonequilibrium stationary state are then derived
in terms of the matrix $\m Q$ or of its associated Hermitian form $\m H$:
stationary probability current (section~\ref{current}),
entropy production rate (section~\ref{entropy}),
correlation, response and fluctuation-dissipation ratio (section~\ref{crfdt}).
Various other facets of the problem are also investigated,
in particular the effect of a linear change of coordinates (section~\ref{reparametrisation}),
a formal investigation of the case where the friction matrix $\m B$
(see~(\ref{mou})) is diagonalisable (section~\ref{biortho}),
cyclically symmetric processes (section~\ref{oucyclic})
and the case of two degrees of freedom (section~\ref{two}).

We then analyse in detail two different examples of physical systems
giving rise to multivariate Ornstein-Uhlenbeck processes.
Our first example (sections~\ref{ring} and~\ref{chain})
is borrowed from the microscopic world of kinetic spin systems.
We consider the one-dimensional ferromagnetic Gaussian spin model
endowed with a stochastic dynamics where spatial asymmetry results in irreversibility,
along the lines of earlier studies on the Ising and spherical ferromagnets~\cite{kun,stau,gb2009,cg2011,oliv,oliverr,cg2013,gp1,GLI,gp2,GLBI,gp3}.
The geometry of a finite ring is discussed in section~\ref{ring}.
In this situation,
the asymmetric dynamics corresponds to a cyclically symmetric process,
and the stationary state is independent of the asymmetry parameter $V$.
We successively investigate the statics of the model (section~\ref{rstat}),
the conventional gradient dynamics (section~\ref{rgradient}),
the asymmetric dynamics (section~\ref{rasymmetric}),
and the key observables characterising the resulting nonequilibrium stationary state,
first in the thermodynamic limit (section~\ref{rthermo})
and then on finite-size systems (section~\ref{rfss}).
Section~\ref{chain} is devoted to the model on a finite open chain.
There, at variance with the ring geometry,
the stationary state depends continuously on the asymmetry parameter $V$.
We study the statics of the model (section~\ref{cstat}),
general properties of the asymmetric dynamics (section~\ref{schain}),
the spectrum of the friction matrix (section~\ref{cspectrum}),
the first few system sizes (section~\ref{fewn}),
and the main features of the nonequilibrium stationary state (section~\ref{cmain}).
We then turn to more detailed investigations of the model,
both in the small-$V$ regime
(section~\ref{smallv}) and in the totally asymmetric case $V=1$ (section~\ref{vone}).
Our second class of examples of multivariate Ornstein-Uhlenbeck processes
originates in macroscopic physics.
We consider arrays of resistively coupled electrical circuits,
where resistors are sources of thermal noise.
Within this setting,
irreversibility is brought about by an inhomogeneous temperature profile.
The main emphasis is on the entropy production rate per unit time
in the regime where the fluctuations of the temperature profile are small.
We successively consider $RL$ networks (section~\ref{rl})
and $RC$ networks (section~\ref{rc}).
The scaling of the entropy production rate
with the network size is very different in both cases.
Section~\ref{discussion} contains a brief discussion of our findings.
Two appendices contain the derivations
of the entropy production amplitude on the open spin chain (\ref{asmallv})
and of the amplitude matrices giving the entropy production rates
on both types of electrical arrays (\ref{aelec}).

\section{General framework}
\label{general}

\subsection{The model}
\label{model}

A multivariate Ornstein-Uhlenbeck process~\cite{WU,L,lax66}
(see also~\cite{TT}, and~\cite{R,VK,G} for reviews)
is a diffusion process defined by $N$ coupled linear Langevin equations of the form
\beq
\frac{\dd x_m(t)}{\dd t}=-\sum_nB_{mn}x_n(t)+\eta_m(t)
\label{ou}
\eeq
($m,n=1,\dots,N$),
i.e., in vector and matrix notations,
\beq
\frac{\dd\m x(t)}{\dd t}=-\m B\m x(t)+\m\eta(t).
\label{mou}
\eeq
The $\eta_m(t)$ are Gaussian white noises such that
\beq
\mean{\eta_m(t)\eta_n(t')}=2D_{mn}\delta(t-t'),
\eeq
i.e.,
\beq
\mean{\m\eta(t)\m\eta^T(t')}=2\m D\delta(t-t').
\eeq
Throughout this work 
boldface symbols denote vectors and matrices,
and the superscript $T$ denotes the transpose of a vector or of a matrix.

The process under study is thus defined by two real matrices of size $N\times N$:
the noise covariance matrix or diffusion matrix $\m D$, which is symmetric by construction,
and the friction matrix $\m B$, which is not symmetric in general.
In what follows, we assume that $\m D$ is a positive definite matrix,
and that~$\m B$ is the opposite of a stability (or Hurwitz) matrix.
This means that all its eigenvalues $\w B_k$, which are in general complex, obey
\beq
\Re\w B_k>0\qquad(k=1,\dots,N).
\eeq
We introduce for further reference the spectral gap
\beq
\w B\imin=\min_k\,\Re\w B_k.
\label{gap}
\eeq
The above assumptions exclude degenerate cases.
They ensure that the process relaxes exponentially fast
to $\m x=\m 0$ in the absence of noise,
and to a fluctuating stationary state with Gaussian statistics in the presence of noise.
The main goal of this work is to characterise various features of the latter state,
which is generically a nonequilibrium stationary state.

\subsection{Relaxation and stationary state}
\label{steady}

The relaxation of the process $\m x(t)$ from its (deterministic) initial value $\m x(0)$
is encoded in the Green's function
\beq
\m G(t)=\e^{-\m B t},
\label{gdef}
\eeq
which obeys
\beq
\frac{\dd\m G(t)}{\dd t}=-\m B\m G(t)=-\m G(t)\m B,
\label{dgdt}
\eeq
with $\m G(0)=\m 1$.
The hypothesis made on the friction matrix $\m B$
ensures that $\m G(t)$ falls off exponentially fast to zero.
We have
\beq
\m x(t)=\m G(t)\m x(0)+\int_0^t\m G(t-s)\m\eta(s)\,\dd s.
\label{xt}
\eeq

The process $\m x(t)$ is Gaussian,
as it can be expressed linearly in terms of the Gaussian white noises $\m\eta(t)$.
Its statistics at any given time $t$ is entirely characterised
by its mean value, $\mean{\m x(t)}=\m G(t)\m x(0)$,
and by its full equal-time correlation matrix
\beq
\m S(t)=\mean{\m x(t)\m x^T(t)}.
\eeq
Equation~(\ref{xt}) implies
\beq
\m S(t)=\m G(t)\m S(0)\m G^T(t)+2\int_0^t\m G(t-s)\m D\m G^T(t-s)\,\dd s.
\label{st}
\eeq
It can be checked that $\m S(t)$ obeys the differential equation
\beq
\frac{\dd\m S(t)}{\dd t}=2\m D-\m B\m S(t)-\m S(t)\m B^T.
\label{dsdt}
\eeq

The stationary state of the process is therefore Gaussian, with zero mean,
and characterised by the stationary covariance matrix
\beq
\m S=\lim_{t\to\infty}\m S(t)
=2\int_0^\infty\m G(t)\m D\m G^T(t)\,\dd t.
\eeq
This stationary state is therefore characterised by the probability density
\beq
P(\m x)=(2\pi)^{-N/2}(\det\m S)^{1/2}\exp\left(-\half\m x^T\m S^{-1}\m x\right).
\label{pstat}
\eeq

Equation~(\ref{dsdt}) implies
\beq
\m B\m S+\m S\m B^T=2\m D.
\label{sid}
\eeq
This is the key equation of the problem,
relating the stationary covariance matrix $\m S$
to the friction and diffusion matrices $\m B$ and $\m D$ defining the process.
Linear matrix equations of this type
are referred to as Sylvester (or Lyapunov) equations.

\subsection{Reversible vs.~irreversible processes}
\label{irreversible}

The condition for the process~(\ref{ou}) to be reversible, known for long~\cite{L,R,VK,G},
is that the matrix product $\m B\m D$ be symmetric, i.e.,
\beq
\m B\m D=\m D\m B^T.
\label{bds}
\eeq
In this situation,
each term in the left-hand side of~(\ref{sid}) is separately equal to $\m D$.
The stationary covariance matrix reads
\beq
\m S=\m B^{-1}\m D=\m D(\m B^T)^{-1}.
\label{srev}
\eeq
The corresponding stationary state is an equilibrium state.

Whenever the symmetry condition~(\ref{bds}) is not obeyed, the process is irreversible,
with a nonequilibrium stationary state.
Solving the Sylvester equation~(\ref{sid}) for the covariance matrix $\m S$
is more difficult than inverting the friction matrix $\m B$.
We parametrise the matrices $\m B$, $\m D$ and $\m S$ by setting
\beq
\m L=\m B\m S=\m D+\m Q,\qquad
\m L^T=\m S\m B^T=\m D-\m Q.
\label{qdef}
\eeq
The matrix $\m L$ is the Onsager matrix of kinetic coefficients~\cite{ELS,G}.
Its antisymmetric part~$\m Q$ provides a measure
of the amount of irreversibility of the process.
If the process is reversible, $\m L=\m D$ is symmetric and $\m Q$ vanishes.

It is advantageous to recast the matrix $\m Q$ in the following balanced form:
\beq
\m Q=\ii\,\m D^{1/2}\m H\m D^{1/2},
\label{hdef}
\eeq
where $\ii$ is the imaginary unit,
whereas $\m D^{1/2}$, the positive square root of the diffusion matrix~$\m D$,
is a symmetric positive definite matrix.
The matrix $\m H$ thus defined is both dimensionless and Hermitian.
Its entries are purely imaginary,
and so $\ii\m H$ is a real antisymmetric matrix.
As a consequence, the eigenvalues $\w H_k$ of $\m H$
occur in pairs of opposite real numbers (unless $\w H_k=0$).
Its spectral radius, i.e., its largest positive eigenvalue,
\beq
\w H\imax=\max_k\,\w H_k,
\label{hmax}
\eeq
is dubbed the asymmetry index of the process~\cite{BS,MBG}.

\subsection{Stationary probability current}
\label{current}

In the stationary state, following the Fokker-Planck approach~\cite{R,VK,G},
the probability current $\m J(\m x)$ reads
\beq
J_m(\m x)=-\sum_n\left(B_{mn}x_n+D_{mn}\frac{\partial}{\partial x_n}\right)P(\m x),
\eeq
where the stationary probability density $P(\m x)$ is given by~(\ref{pstat}).
We thus obtain
\beq
J_m(\m x)=\sum_n\mu_{mn}x_nP(\m x),
\eeq
i.e.,
\beq
\m J(x)=\m\mu\m xP(\m x),
\label{eq:J}
\eeq
where the mobility tensor $\m\mu$ reads (see~(\ref{qdef}))
\beq
\m\mu=\m D\m S^{-1}-\m B=-\m Q\m S^{-1}.
\label{mudef}
\eeq
The stationary probability current is therefore proportional
to $\m Q$, the matrix measuring the irreversibility of the process.

\subsection{Entropy production rate}
\label{entropy}

Associating an increase of entropy to an irreversible process
is one of the possible statements of the second law of Thermodynamics.
In more recent times, much attention has been devoted to the rate
of entropy production per unit time for an open system~\cite{P,GM}.
The latter rate is also the key quantity involved in
the Gallavotti-Cohen fluctuation theorem~\cite{GC}.
By now, it is commonly recognised as being a fundamental quantity
characterising a nonequilibrium stationary state.

In the present context of diffusion processes,
a general expression for the entropy production rate $\Phi$ per unit time
in the stationary state seemingly appears
for the first time in print in~\cite{QQ}
(see also~\cite{HQ2,QZ,HQ1,ZQ,HQ3} and~\cite{TO,LTO}).
In our notations, this reads
\beq
\Phi=\mean{\m x^T(\m D^{-1}\m B-\m S^{-1})^T\m D(\m D^{-1}\m B-\m S^{-1})\m x},
\eeq
where the average is taken over the stationary state measure of the process.
As a consequence of~(\ref{qdef}) and~(\ref{mudef}), we have
$\m D^{-1}\m B-\m S^{-1}=\m D^{-1}\m Q\m S^{-1}=-\m D^{-1}\m\mu$,
hence
\beq
\Phi=-\mean{\m x^T\m S^{-1}\m Q\m D^{-1}\m Q\m S^{-1}\m x}
=\mean{\m x^T\m\mu^T\m D^{-1}\m\mu\m x}.
\eeq
Using (\ref{eq:J}) and (\ref{mudef}),
the rightmost expression can be recast into the more familiar form~\cite{SF}
\beq
\Phi=\int\frac{\m J^T(\m x)\m D^{-1}\m J(\m x)}{P(\m x)}\,\dd\m x.
\eeq
As expected, $\Phi$ is strictly positive if the process is irreversible,
and it vanishes only if the process is reversible.
Furthermore,
since the stationary state is Gaussian with covariance matrix $\m S$,
we have $\mean{\m x^T\m A\m x}=\tr(\m S\m A)$, and so
\beqa
\Phi
&=&-\tr(\m S^{-1}\m Q\m D^{-1}\m Q)
\nonumber\\
&=&\tr(\m S\m\mu^T\m D^{-1}\m\mu)
\nonumber\\
&=&\tr(\m D^{1/2}\m S^{-1}\m D^{1/2}\m H^2).
\label{phi1}
\eeqa
These formulas express the entropy production rate $\Phi$
as a quadratic form in the matrices $\m Q$, $\m\mu$ and $\m H$ characterising the irreversibility of the process.
Finally, using again~(\ref{qdef}),~(\ref{phi1}) can be recast
into the following expressions
\beq
\Phi=\tr(\m B^T\m D^{-1}\m Q)
=-\tr(\m D^{-1}\m B\m Q),
\label{phi2}
\eeq
involving neither the covariance matrix $\m S$ nor its inverse explicitly.

\subsection{Correlation, response and fluctuation-dissipation ratio}
\label{crfdt}

We now turn to an investigation of the dynamics of the process,
both in its transient regime and in its stationary state.
Keeping in line with recent studies of slow dynamics and aging phenomena
(see~\cite{CLZ,LE,C} for reviews),
we focus our attention onto the correlation,
response and fluctuation-dissipation ratio.

The correlation matrix $\m C(t,s)$ is defined as
\beq
C_{mn}(t,s)=\mean{x_m(t)x_n(s)},
\label{cdef}
\eeq
where the two times $s$ and $t$ are such that $0\le s\le t$.
This quantity obeys
\beq
\frac{\partial\m C(t,s)}{\partial t}=-\m B\m C(t,s),
\eeq
with initial value $\m C(s,s)=\m S(s)$ for $t=s$, hence (see~(\ref{gdef}))
\beq
\m C(t,s)=\m G(t-s)\m S(s).
\eeq

The response matrix $\m R(t,s)$ is defined as
\beq
R_{mn}(t,s)=\frac{\delta\mean{x_m(t)}}{\delta h_n(s)},
\label{rdef}
\eeq
where the ordering fields $h_n(t)$ are linearly coupled
to the dynamical variables $x_n(t)$,
i.e., they are added to the noises $\eta_n(t)$ on the right-hand side of~(\ref{ou}).
We thus readily obtain
\beq
\m R(t,s)=\m G(t-s).
\eeq

In the stationary state, keeping the time lag $\tau=t-s\ge0$ fixed,
the stationary correlation and response matrices read
\beq
\m C(\tau)=\m G(\tau)\m S,\qquad\m R(\tau)=\m G(\tau).
\eeq
We have therefore
\beq
\m C(\tau)=\m R(\tau)\m S,\qquad
\frac{\dd\m C(\tau)}{\dd\tau}=-\m R(\tau)\m B\m S.
\label{dcdt}
\eeq
The latter formula can be reshaped by introducing
a dimensionless stationary fluctuation-dissipation ratio (FDR) matrix $\m X$, such that
\beq
\m R(\tau)\m D=-\frac{\dd\m C(\tau)}{\dd\tau}\m X.
\label{xdef}
\eeq
This definition agrees with the usage in the literature
on slow dynamics and aging phenomena~\cite{C,cuku}.
The diffusion matrix $\m D$ plays the role of temperature.
The fact that $\m D$ and $\m X$ occur in the rightmost positions
on both sides of~(\ref{xdef}) is due
to the conventions used in the definitions~(\ref{cdef}),~(\ref{rdef})
of the correlation and response matrices.
We thus obtain
\beq
\m X^{-1}=\m D^{-1}\m B\m S.
\eeq
Using~(\ref{qdef}) and~(\ref{hdef}) yields
\beq
\m X^{-1}=\m 1+\m D^{-1}\m Q=\m 1+\ii\,\m D^{-1/2}\m H\m D^{1/2}.
\label{xres}
\eeq

\begin{figure}[!ht]
\begin{center}
\includegraphics[angle=0,width=.5\linewidth]{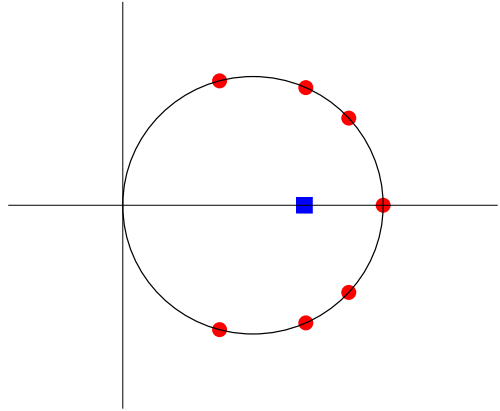}
\caption{\small
A sketch of the spectrum of the FDR matrix $\m X$
in a typical situation with $N=7$.
Red circles: individual eigenvalues $\w X_k$.
Blue square: their center of mass $X\typ$.}
\label{spx}
\end{center}
\end{figure}

In the case of a reversible process ($\m Q=\m 0$), we have $\m X=\m 1$
and the equilibrium fluctuation-dissipation theorem is recovered as
\beq
\m R(\tau)\m D = -\frac{\dd\m C(\tau)}{\dd\tau}.
\eeq
In the general situation of an irreversible process,
the FDR matrix $\m X$ is non-trivial.
The second expression of~(\ref{xres}) shows that its eigenvalues are of the form
\beq
\w X_k=\frac{1}{1+\ii\w H_k},
\eeq
where $\w H_k$ are the eigenvalues of $\m H$, introduced below~(\ref{hdef}).
The $\w X_k$ therefore lie on the circle with diameter [0,1] in the complex plane,
and they occur in complex conjugate pairs (unless $\w X_k=1$).
The center of mass of these eigenvalues defines the typical FDR
\beq
X\typ=\frac{1}{N}\tr\m X=\frac{1}{N}\sum_k\frac{1}{1+\w H_k^2}.
\label{xtyp}
\eeq
This quantity is real and such that $0<X\typ\le1$,
the upper bound $X\typ=1$ being attained for reversible processes.
Figure~\ref{spx} shows a sketch of the spectrum of the FDR matrix $\m X$
in a typical situation with $N=7$.

\subsection{Considerations on reparametrisation}
\label{reparametrisation}

In this section we investigate the effect of a reparametrisation,
i.e., of a change of coordinates.
In order to keep the linearity of the process~(\ref{ou}),
we restrict ourselves to a linear change of coordinates of the form
\beq
\m y=\m M\m x,
\label{re}
\eeq
where $\m M$ is an invertible matrix.

In terms of the new coordinates $\m y$, the process is transformed into a similar one,
characterised by the new matrices
\beq
\m B\new=\m M\m B\m M^{-1},\qquad
\m D\new=\m M\m D\m M^T.
\label{bdnew}
\eeq
The matrices $\m S$ and $\m Q$ transform as $\m D$, i.e.,
\beq
\m S\new=\m M\m S\m M^T,\qquad
\m Q\new=\m M\m Q\m M^T,
\eeq
whereas the matrices $\m H$ and $\m X$ transform as
\beq
\m H\new=\m W^{-1}\m H\m W,\qquad
\m W=\m D^{1/2}\m M^T(\m M\m D\m M^T)^{-1/2},
\eeq
and
\beq
\m X\new=(\m M^T)^{-1}\m X\m M^T.
\eeq

The matrices $\m B\new$, $\m H\new$ and $\m X\new$ are respectively conjugate
to $\m B$, $\m H$ and $\m X$,
so that their spectra are invariant under reparametrisation.
The entropy production rate~$\Phi$, the typical FDR $X\typ$
and the asymmetry index $\w H\imax$ are also invariant under reparametrisation.
This corroborates the intrinsic nature of these quantities.

It is worth investigating to what extent the process~(\ref{ou})
can be simplified by means of a suitably chosen linear reparametrisation.
Not much can be gained on the side of the friction matrix,
as $\m B$ and $\m B\new$ have the same spectrum.
This invariance could be expected as well,
as the spectrum of $\m B$ encodes the relaxation times of the process.
The diffusion matrix $\m D$ can however, at least in principle,
always be brought to the canonical form
\beq
\m D\new=\m 1,
\label{dcan}
\eeq
corresponding to independent normalised white noises,
by choosing $\m M=\m D^{-1/2}$.

\subsection{The case where the friction matrix $\m B$ is diagonalisable}
\label{biortho}

As already stated,
solving the Sylvester equation~(\ref{sid}) for the covariance matrix $\m S$
is more difficult than inverting the friction matrix $\m B$.
The generic situation where $\m B$ is diagonalisable
can however be formally dealt with as follows~\cite{R}.
Assume that $\m B$ has a biorthogonal system of left and right eigenvectors
such that $\bra{\ell_k}\m B=\w B_k\bra{\ell_k}$ and $\m B\ket{r_k}=\w B_k\ket{r_k}$,
in Dirac notation, with
\beq
\braket{\ell_k}{r_l}=\delta_{kl},\qquad
\sum_k\ket{r_k}\bra{\ell_k}=\m 1.
\eeq
We have then
\beq
\m G(t)=\sum_k\e^{-\w B_kt}\ket{r_k}\bra{\ell_k}.
\eeq
Equation~(\ref{st}) implies
\beqa
\m S(t)
&=&\sum_{kl}\e^{-(\w B_k+\w B_l)t}
\ket{r_k}\braopket{\ell_k}{\m S(0)}{\ell_l}\bra{r_l}
\nonumber\\
&+&2\sum_{kl}\frac{1-\e^{-(\w B_k+\w B_l)t}}{\w B_k+\w B_l}
\,\ket{r_k}\braopket{\ell_k}{\m D}{\ell_l}\bra{r_l}.
\eeqa
In particular, the stationary covariance matrix reads
\beq
\m S=2\sum_{kl}
\frac{\ket{r_k}\braopket{\ell_k}{\m D}{\ell_l}\bra{r_l}}{\w B_k+\w B_l}.
\label{ssp}
\eeq
After reduction the denominator of this expression reads
\beq
\Delta_N=\prod_k\w B_k\,\prod_{k<l}(\w B_k+\w B_l).
\label{pden}
\eeq
The first factor, equal to $\det\m B$, has degree $N$ in the entries of $\m B$,
whereas the second factor has degree $N(N-1)/2$.
Altogether, $\Delta_N$ is a polynomial of degree $N(N+1)/2$.
This degree equals the number of independent entries in the symmetric matrix $\m S$.

With the same conventions, we have
\beq
\m Q=\sum_{kl}\frac{\w B_k-\w B_l}{\w B_k+\w B_l}
\,\ket{r_k}\braopket{\ell_k}{\m D}{\ell_l}\bra{r_l}.
\label{qsp}
\eeq
When the process is reversible, i.e., if~(\ref{bds}) holds,
it can be checked that the matrix element $\braopket{\ell_k}{\m D}{\ell_l}$
vanishes whenever $\w B_k$ and $\w B_l$ are different,
and so $\m Q$ vanishes, as should be.

This scheme will be used several times in the course of this work,
starting with section~\ref{oucyclic}.
In the happy instances where either $\m B$ and $\m D$ commute or $\m D$ is simple enough,
quantities of physical interest such as $\Phi$
can also be expressed in terms of the eigenvectors of $\m B$.

\subsection{Cyclic symmetry}
\label{oucyclic}

As usual, the complexity of the problem can be reduced in the presence of symmetries.
In this section we consider cyclically symmetric processes,
where the dynamical variables live on the sites labelled $n$ of a ring of~$N$ points
and the dynamics is invariant under discrete translations along the ring.
The matrices $\m B$ and $\m D$ are therefore circulant matrices,
whose entries $B_{mn}=B_{m-n}$ and $D_{mn}=D_{m-n}$ only depend on the difference
$m-n$~mod~$N$.
For instance, for $N=4$ this reads
\beq
\m B=\pmatrix{B_0&B_1&B_2&B_3\cr B_3&B_0&B_1&B_2\cr B_2&B_3&B_0&B_1\cr B_1&B_2&B_3&B_0},
\quad
\m D=\pmatrix{D_0&D_1&D_2&D_3\cr D_3&D_0&D_1&D_2\cr D_2&D_3&D_0&D_1\cr D_1&D_2&D_3&D_0}.
\eeq
The property that $\m D$ is symmetric imposes $D_1=D_3$.

In this setting, it is natural to use the discrete Fourier transform
\beq
\w x_k=\sum_n\e^{-2\pi\ii kn/N}x_n,\qquad
x_n=\frac{1}{N}\sum_k\e^{2\pi\ii kn/N}\w x_k.
\eeq
We recall that the indices $n$ and $k$ are understood mod~$N$.
If needed, they can therefore be restricted to the range $n,k=1,\dots,N$.
Cyclic symmetry brings out a considerable simplification.
All matrices pertaining to the problem are diagonal in the Fourier basis,
and therefore commute with each other.
The eigenvalues of a cyclically symmetric matrix $\m A$
indeed coincide with its discrete Fourier amplitudes $\w A_k$.

The diffusion matrix $\m D$ is symmetric,
i.e., we have $D_n=D_{-n}$, and so $\w D_k=\w D_{-k}$ is real.
The friction matrix $\m B$ is not symmetric in general,
and so $\w B_k$ is complex, and $(\w{B^T})_k=\w B_{-k}$ is its complex conjugate,
so that
\beq
\w B_k+(\w{B^T})_k=2\,\Re\w B_k,\qquad
\w B_k-(\w{B^T})_k=2\ii\,\Im\w B_k.
\eeq
In this context, the process is reversible if and only $\m B$ is symmetric,
i.e., $\w B_k=\w B_{-k}$ is real.
The Sylvester equation~(\ref{sid}) yields
\beq
\w S_k=\frac{\w D_k}{\Re\w B_k}.
\label{sk}
\eeq
We have thus
\beqa
\w H_k&=&\frac{\Im\w B_k}{\Re\w B_k},
\label{hk}
\\
\w Q_k&=&\ii\w H_k\w D_k,
\\
\w X_k&=&\frac{1}{1+\ii\w H_k}=\frac{\Re\w B_k}{\w B_k},
\label{xk}
\\
\w\mu_k&=&-\ii\,\Im\w B_k,
\eeqa
and finally
\beq
\Phi=\sum_k\frac{(\Im\w B_k)^2}{\Re\w B_k}.
\label{phik}
\eeq

Let us add a remark on finite-ranged matrices.
Let $\m B$ be a circulant matrix such that $B_n\ne0$ for $\abs{n}\le r$ only.
The integer $r$ is dubbed the range of the matrix $\m B$.
As soon as the system size obeys $N>2r$,
the amplitudes $\w B_k$ labelled by the discrete Fourier index $k$
are the restriction to the quantised momenta
\beq
q_k=\frac{2\pi k}{N}\qquad(k=1,\dots,N)
\eeq
of the trigonometric polynomial
\beq
\w B(q)=\sum_n\e^{-\ii nq}B_n,
\eeq
which is independent of $N$.
When $\m B$ is symmetric, the above condition on the system size can be relaxed to $N\ge 2r$.
We shall see an example of such a situation with $r=1$ in section~\ref{ring},
devoted to the Gaussian spin model on a ring.

\subsection{The case $N=2$}
\label{two}

To close this general section,
we give explicit expressions of all quantities introduced so far
in the case of two coupled degrees of freedom.
This situation already exhibits most of the generic features
of multivariate Ornstein-Uhlenbeck processes
(see~\cite{BaMN} for a recent account).

We parametrise the friction and diffusion matrices as
\beq
\m B=\pmatrix{a&b\cr c&d},\qquad
\m D=\pmatrix{u&w\cr w&v}.
\eeq
The friction matrix $\m B$ has positive eigenvalues for $a+d>0$ and $\delta=ad-bc>0$,
whereas the diffusion matrix $\m D$ is positive definite for $u>0$ and $uv-w^2>0$.

The reversibility condition~(\ref{bds}) amounts to one single bilinear condition:
\beq
cu-bv+(d-a)w=0.
\label{s2}
\eeq
The solution of the Sylvester equation~(\ref{sid}) reads
\beqa
\m S&=&\frac{1}{(a+d)\delta}
\nonumber\\
&\times&\pmatrix{(\delta+d^2)u+b^2v-2bd w&-cd u-ab v+2ad w\cr
-cd u-ab v+2ad w&c^2u+(\delta+a^2)v-2ac w}.
\eeqa

Let us introduce the irreversibility parameter (see~(\ref{s2}))
\beq
g=\frac{cu-bv+(d-a)w}{(a+d)(uv-w^2)}.
\eeq
Equation~(\ref{qdef}) yields
\beq
\m Q=(uv-w^2)g\pmatrix{0&-1\cr 1&0}.
\eeq
The entropy production rate reads
\beq
\Phi=(a+d)(uv-w^2)g^2=\frac{(cu-bv+(d-a)w)^2}{(a+d)(uv-w^2)}.
\eeq
Equation~(\ref{xres}) yields
\beq
\m X^{-1}=\pmatrix{1-wg&-vg\cr ug&1+wg}
\eeq
and
\beq
\m X=\frac{1}{1+(uv-w^2)g^2}\pmatrix{1+wg&vg\cr -ug&1-wg}.
\eeq
We have therefore
\beq
X\typ=\frac{1}{1+(uv-w^2)g^2}
\eeq
and
\beq
\w H\imax=\sqrt{uv-w^2}\,\abs{g}.
\eeq

In the present situation,
there is one single reversibility condition~(\ref{s2}),
and therefore one single irreversibility parameter $g$.
As a consequence,
all the quantities characterising the irreversibility are related to each other.
We have indeed
\beq
\Phi=(a+d)\w H\imax^2,\qquad X\typ=\frac{1}{1+\w H\imax^2}.
\eeq

For higher values of $N$,
the reversibility condition~(\ref{bds}) amounts to $N(N-1)/2$ conditions.
The complexity of the expressions of the various quantities of interest
grows very fast with $N$ (see below~(\ref{pden})).

\section{The Gaussian spin model on a ring}
\label{ring}

Our first example of a multivariate Ornstein-Uhlenbeck process
originates in a microscopic model,
which is the one-dimensional version of the ferromagnetic Gaussian spin model~\cite{berlin}.
We shall endow the model with a dynamics
where spatial asymmetry results in irreversibility.
The main control parameters will be the system size $N$
and the asymmetry parameter $V$.
In this section we investigate the dynamics on a ring,
whereas the more intricate situation of an open chain
will be dealt with in section~\ref{chain}.

\subsection{Statics}
\label{rstat}

Let us begin with a brief report on the statics of the model.
We consider the geometry of a finite ring of $N$ sites ($N\ge2$), with periodic boundary conditions.
Each site~$n$ hosts a continuous spin $x_n$.
Throughout section~\ref{ring}, the index $n$ is understood mod~$N$.
At infinite temperature,
each spin has a Gaussian distribution such that $\mean{x_n^2}=1$.
The weight of a spin configuration is therefore proportional to $\exp(-\A_0)$,
where the free action $\A_0$ reads
\beq
\A_0=\frac{1}{2}\sum_nx_n^2.
\eeq
Interactions between spins are described by the nearest-neighbor ferromagnetic Hamiltonian
\beq
\H=-J\sum_nx_nx_{n+1},
\eeq
with $J>0$.
At temperature $T=1/\beta$,
the weight of a spin configuration is therefore proportional to $\exp(-\A)$,
where the full action reads
\beq
\A=\A_0+\beta\H=\frac{1}{2}\sum_n(x_n^2-2Kx_nx_{n+1}),
\label{afull}
\eeq
with
\beq
K=\beta J.
\eeq

The model owes its name to the fact that its Boltzmann weight is Gaussian.
Its statics is characterised by the covariance matrix $\m S$
whose entries are $S_{mn}=\mean{x_mx_n}$.
By definition,
$\m S$ is the inverse of the symmetric matrix~$\m A$ associated
with the quadratic form~(\ref{afull}).
The non-zero elements of $\m A$ are $A_{nn}=1$ and $A_{n,n\pm1}=-K$,
with periodic boundary conditions.
Translational invariance allows one to use the discrete Fourier transform
introduced in section~\ref{oucyclic}.
As anticipated there,
all the quantities labelled by the Fourier index $k$
are the restrictions of smooth functions of $q$ to the discrete set of momenta
\beq
q_k=\frac{2\pi k}{N}\qquad(k=1,\dots,N).
\eeq
In Fourier space, with the same conventions as in section~\ref{oucyclic}, we have
\beq
\w A(q)=1-2K\cos q,\qquad\w S(q)=\frac{1}{1-2K\cos q},
\label{sstat}
\eeq
and therefore $S_{mn}=S_{m-n}$, with
\beq
S_n=\frac{1}{N}\sum_k\frac{\e^{\ii nq_k}}{1-2K\cos q_k}.
\label{snsum}
\eeq
The range of $\m A$ is $r=1$, and so the above expression holds for all $N\ge2$.

The model is well-defined as long as the quadratic form~(\ref{afull})
is positive definite.
This condition amounts to requesting that all the eigenvalues $\w A_k$,
which appear as denominators in~(\ref{snsum}), are positive.
The smallest of them, corresponding to $k=N$, i.e., $q_k=0$, reads $\w A_N=1-2K$.
It vanishes at the critical point
\beq
K_c=\frac{1}{2}.
\eeq
The model therefore exists only in its high-temperature phase ($K<K_c$).
The existence of a non-zero critical temperature in one dimension
underlines the lack of realism of the ferromagnetic Gaussian model~\cite{berlin}.
In spite of this, the dynamics of this model provides interesting examples
within the framework of the present study.

In order to derive a closed-form expression of the correlation function $S_n$,
instead of evaluating the sum in~(\ref{snsum}),
it is more convenient to observe that it obeys the difference equation
\beq
S_n=K(S_{n+1}+S_{n-1})+\delta_{n0},
\eeq
with periodic boundary conditions, expressing the identity $\m A\m S=\m 1$.
Looking for a solution of the form $S_n=A\,\e^{n\mu}+B\,\e^{-n\mu}$,
we obtain after some algebra
\beq
S_n=\cotanh\mu\,\frac{\cosh\left(n-\frac{N}{2}\right)\mu}{\sinh\frac{N\mu}{2}}
\qquad(n=0,\dots,N).
\label{snring}
\eeq
The mass (inverse correlation length) $\mu$ is given by
\beq
\cosh\mu=\frac{1}{2K},
\eeq
and vanishes at the critical point according to
\beq
K_c-K\approx\frac{\mu^2}{4}.
\eeq
The correlation function diverges uniformly as
\beq
S_n\approx\frac{2}{N\mu^2},
\label{rcrit}
\eeq
irrespective of the distance $n$, as the critical point is approached.

In the thermodynamic limit ($N\to\infty$), keeping $n$ fixed,
the expression~(\ref{snring}) simplifies to
\beq
S_n=\cotanh\mu\,\e^{-\abs{n}\mu}.
\label{snbulk}
\eeq

\subsection{Gradient dynamics}
\label{rgradient}

The most natural dynamics for the Gaussian spin model
is the gradient dynamics defined by
\beq
\frac{\dd x_n(t)}{\dd t}=-\frac{\partial\A}{\partial x_n}+\eta_n(t)
=-x_n+K(x_{n-1}+x_{n+1})+\eta_n(t),
\label{ssym}
\eeq
with periodic boundary conditions.
The $\eta_n(t)$ are independent white noises such that
\beq
\mean{\eta_m(t)\eta_n(t')}=2\delta_{mn}\delta(t-t').
\eeq
This dynamics corresponds to an Ornstein-Uhlenbeck process
with cyclic symmetry, as defined in section~\ref{oucyclic}.
The diffusion matrix $\m D$ is the unit matrix,
whereas the friction matrix $\m B$ coincides with the matrix $\m A$
associated with the quadratic form~(\ref{afull}).
The proportionality between $\m A$ and $\m B$ is a common characteristic feature
of linear gradient dynamics.
In particular, the matrix $\m B$ is symmetric, and so the process is reversible.
The expression $\m S=\m B^{-1}$ of the covariance matrix (see~(\ref{srev}))
coincides with the static result $\m S=\m A^{-1}$.
We thus readily recover the static properties of the model recalled in section~\ref{rstat}.

\subsection{Asymmetric dynamics}
\label{rasymmetric}

We now deform the gradient dynamics (\ref{ssym}) into the following asymmetric one:
\beq
\frac{\dd x_n(t)}{\dd t}=-x_n+K((1+V)x_{n-1}+(1-V)x_{n+1})+\eta_n(t),
\label{sasym}
\eeq
with an arbitrary spatial asymmetry parameter ($-\infty<V<+\infty$).
This is the most general form keeping both the linearity and the range of
dynamical interactions.
The ferromagnetic spherical model with an asymmetric dynamics of this kind
has been investigated in~\cite{HO}, and in more detail in~\cite{GLSA}.
Ising spin models where spatially asymmetric rules
induce irreversibility have been studied
as well~\cite{maes,kun,stau,gb2009,cg2011,oliv,oliverr,cg2013,gp1,GLI,gp2,GLBI,gp3}.

The asymmetric dynamics~(\ref{sasym})
again corresponds to an Ornstein-Uhlenbeck process with cyclic symmetry,
as defined in section~\ref{oucyclic}.
The diffusion matrix is still the unit matrix.
The non-zero elements of the friction matrix are $B_0=1$,
$B_1=-K(1-V)$ and $B_{-1}=-K(1+V)$.

The irreversibility of the process is measured
by the spatial asymmetry para\-meter~$V$:
as soon as $V$ is non-zero, $\m B$ is not symmetric,
and the process is irreversible.
The remainder of this section~\ref{ring} is devoted
to characterising the non\-equilibrium stationary state
associated with the asymmetric dynamics~(\ref{sasym}) on a ring.
In Fourier space, with the same conventions as in section~\ref{oucyclic}, we have
$\w D(q)=1$, whereas
\beq
\w B(q)=1-2K(\cos q+\ii V\sin q).
\label{bq}
\eeq
The eigenvalues $\w B_k=\w B(q_k)$ of $\m B$ are therefore complex.
They lie on the ellipse centered at unity with semi-axes $2K$ and $2K\abs{V}$.
Their real parts $\Re\w B_k$ do not depend on the asymmetry parameter $V$.
In particular, the spectral gap (see~(\ref{gap}))
\beq
\w B\imin=1-2K\approx\frac{\mu^2}{2}.
\label{gapring}
\eeq
is independent of $V$ and vanishes as the critical point is approached.

Using~(\ref{sk}), we obtain an expression of $\w S(q)$
which coincides with~(\ref{sstat}), irrespective of $V$.
In other words, the stationary state is independent of the asymmetry parameter $V$,
i.e., of the irreversibility of the process.
This property originates in the process being cyclically symmetric.
The very same property was already observed in the spherical model
with spatially asymmetric dynamics in the thermodynamic limit
in any dimension~\cite{GLSA}.
It is however not granted in general.
It will indeed turn out to be violated in the geometry of an open chain
(see section~\ref{schain}).

\subsection{Thermodynamic limit}
\label{rthermo}

Within the formalism exposed in section~\ref{oucyclic},
the expression~(\ref{bq}) yields explicit results for intensive quantities
characterising the nonequilibrium stationary state
in the thermodynamic limit ($N\to\infty$),
where sums over $k$ become integrals over $q$.

As a first example, we consider the central moments of the spectrum of $\m B$, i.e.,
\beq
\tau_m=\tr(\m B-\m 1)^{2m}.
\eeq
In the thermodynamic limit, we have
\beqa
\lim_{N\to\infty}\frac{\tau_m}{N}
&=&(2K)^{2m}\int_0^{2\pi}\frac{\dd q}{2\pi}\,(\cos q+\ii V\sin q)^{2m}
\nonumber\\
&=&\frac{(2m)!}{m!^2}K^{2m}(1-V^2)^m.
\label{taures}
\eeqa

Let us now turn to physical quantities.
The entropy production rate is extensive,
i.e., there is a well-defined intensive entropy production rate per unit time and per spin,
\beq
\varphi=\lim_{N\to\infty}\frac{\Phi}{N},
\eeq
which reads
\beqa
\varphi&=&4K^2V^2\int_0^{2\pi}\frac{\dd q}{2\pi}\,\frac{\sin^2q}{1-2K\cos q}
\nonumber\\
&=&\left(1-\sqrt{1-4K^2}\right)V^2=(1-\tanh\mu)V^2.
\label{philim}
\eeqa
This quantity is proportional to $V^2$,
with an amplitude which remains finite at the critical point.

We have furthermore
\beq
\w H(q)=-\frac{2KV\sin q}{1-2K\cos q}.
\label{hq}
\eeq
The asymmetry index of the process (see~(\ref{hmax})) is reached for $\cos q=2K$ and reads
\beq
\w H\imax=\frac{2K\abs{V}}{\sqrt{1-4K^2}}=\frac{\abs{V}}{\sinh\mu}.
\label{hlim}
\eeq
This result is proportional to $\abs{V}$,
with an amplitude which diverges as the critical point is approached.

Finally, the typical FDR reads~(see~(\ref{xk}))
\beqa
X\typ&=&\int_0^{2\pi}\frac{\dd q}{2\pi}\,
\frac{(1-2K\cos q)^2}{(1-2K\cos q)^2+4K^2V^2\sin^2q}
\nonumber\\
&=&\frac{1}{1-V^2}\left(1-\frac{V^2}{\sqrt{1-4K^2(1-V^2)}}\right)
\nonumber\\
&=&\frac{1}{1-V^2}\left(1-\frac{V^2\cosh\mu}{\sqrt{V^2+\sinh^2\mu}}\right).
\label{xlim}
\eeqa
This quantity exhibits a richer dependence on parameters.
For $V\to0$, it departs from its equilibrium value $X\typ=1$ as
\beq
X\typ\approx1-\left(\frac{1}{\sqrt{1-4K^2}}-1\right)V^2\approx1-(\cotanh\mu-1)V^2.
\eeq
For $V=\pm1$, i.e., when interactions are totally asymmetric, we have $X\typ=1-2K^2$.
For $V\to\pm\infty$, $X\typ$ falls off as
\beq
X\typ\approx\frac{1}{2K\abs{V}}\approx\frac{\cosh\mu}{\abs{V}}.
\eeq
At the critical point,~(\ref{xlim}) simplifies to
\beq
X\typ=\frac{1}{1+\abs{V}}.
\eeq

\subsection{Finite-size effects}
\label{rfss}

Intensive quantities pertaining to finite-size samples
generically converge exponentially fast to their thermodynamic limit,
except near the critical point, where slow convergence properties
and finite-size scaling laws can be expected.
Let us take the example of the entropy production rate $\Phi$,
which is
the easiest to analyse from a technical viewpoint.
Equations~(\ref{phik}) and~(\ref{bq}) yield for all $N\ge2$
\beq
\Phi=4K^2V^2\sum_k\frac{\sin^2q_k}{1-2K\cos q_k}.
\eeq
The sum can be performed exactly by means of the identity
(see e.g.~\cite[Eq.~(41.2.8)]{hansen})
\beq
\sum_{k=1}^N\frac{\sinh\mu}{\cosh\mu-\cos q_k}=N\,\cotanh\frac{N\mu}{2}.
\eeq
We obtain after some algebra the closed-form expression
\beq
\Phi=N\left(1-\tanh\mu\,\cotanh\frac{N\mu}{2}\right)V^2.
\eeq
For $N=2$, the parameter $V$ drops out of the dynamical equations~(\ref{sasym}),
so that the process is symmetric and reversible, and so $\Phi$ vanishes.

All over the high-temperature phase,
the difference $\Phi-N\varphi$ falls off as $\e^{-N\mu}$.
Right at the critical point, we have
\beq
\Phi=(N-2)V^2,
\eeq
with a non-vanishing finite-size correction,
a rather unusual feature for a system with periodic boundary conditions.
Finally, in the scaling region around the critical point,
i.e., for $N$ large and $\mu$ small,
the entropy production rate obeys a finite-size scaling law of the form
\beq
\Phi-NV^2\approx-N\mu\,\cotanh\frac{N\mu}{2}\,V^2.
\eeq
The product $N\mu$ is the natural scaling variable of the problem,
as it is the dimensionless ratio between the system size $N$
and the correlation length $1/\mu$.

\section{The Gaussian spin model on an open chain}
\label{chain}

In this section we investigate the dynamics
of the ferromagnetic Gaussian spin model on an open chain.

\subsection{Statics}
\label{cstat}

Let us start with the statics of the model.
We consider the geometry of a finite open chain of $N$ sites ($n=1,\dots,N$), where
each site $n$ hosts a continuous spin~$x_n$.
At temperature $T=1/\beta$,
the weight of a spin configuration is still proportional to $\exp(-\A)$,
with the full action $\A$ given by~(\ref{afull}), albeit with Dirichlet boundary conditions
($x_0=x_{N+1}=0$).
The matrix $\m S$,
whose entries are the correlation functions $S_{mn}=\mean{x_mx_n}$,
is again the inverse of the symmetric matrix~$\m A$ associated with
the latter quadratic form.
We thus have
\beq
S_{mn}=K(S_{m,n+1}+S_{m,n-1})+\delta_{mn},
\eeq
with Dirichlet boundary conditions in both variables
($S_{0n}=S_{N+1,n}=S_{m0}=S_{m,N+1}=0$).
Looking for solutions of the form $S_{mn}=A\,\e^{n\mu}+B\,\e^{-n\mu}$,
separately for $m\le n$ and for $m\ge n$,
we obtain after some algebra the expression
\beq
S_{mn}=\cotanh\mu\,\frac{2\sinh m\mu\,\sinh(N+1-n)\mu}{\sinh(N+1)\mu}\qquad(m\le n),
\label{smnsym}
\eeq
to be completed by symmetry for $m\ge n$.
The correlation function takes finite values at the critical point, i.e.,
\beq
S_{mn}=\frac{2m(N+1-n)}{N+1}\qquad(m\le n).
\label{ccrit}
\eeq
whereas it diverges according to~(\ref{rcrit}) in the ring geometry.
On an open chain, Dirichlet boundary conditions indeed prevent this divergence.
In the thermodynamic limit ($N\to\infty$), keeping the difference $n-m$ fixed,
this expression simplifies to
\beq
S_{mn}=\cotanh\mu\,\e^{-\abs{n-m}\mu},
\label{smnbulk}
\eeq
in agreement with~(\ref{snbulk}).

\subsection{Dynamics}
\label{schain}

We endow the Gaussian model on an open chain
with an asymmetric dynamics along the lines of~(\ref{sasym}), i.e.,
\beq
\frac{\dd x_n(t)}{\dd t}=-x_n+K((1+V)x_{n-1}+(1-V)x_{n+1})+\eta_n(t),
\label{scasym}
\eeq
with Dirichlet boundary conditions, and an arbitrary asymmetry parameter $V$.

The diffusion matrix is again the unit matrix ($\m D=\m 1$).
The non-zero elements of the friction matrix read
\beqa
B_{nn}=1,
\nonumber\\
B_{n,n-1}=-K(1+V)\qquad(n\ne1),
\nonumber\\
B_{n,n+1}=-K(1-V)\qquad(n\ne N).
\label{belts}
\eeqa
In other words, $\m B$ is a tridiagonal Toeplitz matrix.
The irreversibility of the process is again measured
by the spatial asymmetry parameter $V$:
as soon as $V$ is non-zero,~$\m B$ is not symmetric and the process is irreversible.

The remainder of this section~\ref{chain} is devoted
to characterising the nonequilibrium stationary state
of the process defined by~(\ref{scasym}).
The main novel feature with respect to the ring geometry
is that the stationary state now depends on the asymmetry parameter $V$.
Thus, as already stated, deforming the dynamics by $V$
leaves the stationary state unchanged in the ring geometry,
but changes it continuously in the geometry of an open chain.
In the present situation, the Sylvester equation~(\ref{sid}),~i.e.,
\beqa
2S_{mn}&=&K((1+V)(S_{m-1,n}+S_{m,n-1})+(1-V)(S_{m+1,n}+S_{m,n+1}))
\nonumber\\
&+&2\delta_{mn},
\label{sexpl}
\eeqa
with Dirichlet boundary conditions, must be solved numerically in general.
It can be dealt with analytically in some special circumstances only,
namely $N$ small, $V\to0$ and $V=\pm1$ (see respectively sections~\ref{fewn},
\ref{smallv} and~\ref{vone}).

\subsection{The spectrum of $\m B$}
\label{cspectrum}

We begin with the analysis of the spectrum of the friction matrix $\m B$.
Let~$\ket{r_k}$ be a right eigenvector of~$\m B$, with eigenvalue $\w B_k$.
Its components $r_{k,n}=\braket{n}{r_k}$ obey
\beq
(1-\w B_k)r_{k,n}-K(1+V)r_{k,n-1}-K(1-V)r_{k,n+1}=0.
\label{eqr}
\eeq
Assuming for a while $\abs{V}<1$, let us set
\beq
r_{k,n}=\left(\frac{1+V}{1-V}\right)^{n/2}\psi_{k,n}.
\label{eqrphi}
\eeq
This brings~(\ref{eqr}) to the more familiar symmetric form
\beq
(1-\w B_k)\psi_{k,n}-K\sqrt{1-V^2}(\psi_{k,n-1}+\psi_{k,n+1})=0.
\eeq
An orthonormal basis of eigenvectors reads
\beq
\psi_{k,n}=\sqrt{\frac{2}{N+1}}\sin nq_k.
\label{psidef}
\eeq
Throughout section~\ref{chain},
the quantised momenta fit to Dirichlet boundary conditions read
\beq
q_k=\frac{k\pi}{N+1}\qquad(k=1,\dots,N).
\label{qdir}
\eeq
The corresponding eigenvalues are
\beq
\w B_k=1-2K\sqrt{1-V^2}\cos q_k.
\label{eqb}
\eeq
The above expressions can be analytically continued to the range $\abs{V}>1$,
where the square roots entering~(\ref{eqrphi}) and~(\ref{eqb}) become imaginary.

The spectrum of $\m B$ on an open chain is very different from that on a ring.
The eigenvalues $\w B_k$ are real for $\abs{V}<1$,
whereas for $\abs{V}>1$ they have a constant real part
equal to unity and variable imaginary parts.
The spectral gap reads (see~(\ref{gap}))
\beq
\w B\imin=\left\{\matrix{
1-2K\sqrt{1-V^2}\cos\frad{\pi}{N+1}\quad&(\abs{V}<1),\cr
1\hfill&(\abs{V}>1).}\right.
\label{gapchain}
\eeq
The gap is therefore an increasing function of $\abs{V}$ as long as $\abs{V}<1$,
which then saturates to a constant.
This reflects a very general fact.
Irreversibility is known to
generically accelerate the dynamics of diffusions and other classes
of stochastic processes~\cite{HHS1,HHS2,LN,WH,jack}.
This feature however does not hold when the process is cyclically symmetric
(see section~\ref{ring}).
Note that, even in the presence of a cyclic symmetry,
dynamics is accelerated by irreversibility
for discrete spin models~\cite{gb2009,cg2011,gp1}.

In the scaling region near the critical point
where $\mu$ and~$V$ are small, whereas $N$ is large,
the expression~(\ref{gapchain}) simplifies to the sum of three contributions:
\beq
\w B\imin\approx\frac{1}{2}\left(\mu^2+V^2+\frac{\pi^2}{N^2}\right),
\eeq
while only the first one was present in the ring geometry~(see~(\ref{gapring})).

In spite of the differences underlined above,
the spectra of $\m B$ on a ring and on an open chain
in the $N\to\infty$ limit
share the same central moments $\tau_m$ given by~(\ref{taures}).
In the borderline cases $V=\pm1$, interactions become totally asymmetric.
The friction matrix $\m B$ for an open chain is either upper or lower triangular.
Its spectrum consists of the single eigenvalue $\w B=1$ with maximal multiplicity $N$,
whereas that on a ring consists of $N$ equally spaced points on a circle
of radius $2K$ centered at unity.
Both above spectra have vanishing central moments, for all $N\ge2$.

\subsection{The first few values of $N$}
\label{fewn}

It is worth having a closer look at the first few values of $N$.

\subsubsection*{$N=2$.}

The problem is already non-trivial, in contrast with the case of a ring.
The Sylvester equation~(\ref{sid}) yields
\beq
\m S=\frac{1}{1-K^2(1-V^2)}\pmatrix{1-K^2V(1-V)& K\cr K&1+K^2V(1+V)}.
\eeq
The entropy production rate reads
\beq
\Phi=2K^2V^2.
\label{phin2}
\eeq

\subsubsection*{$N=3$.}

The Sylvester equation~(\ref{sid}) yields
\beq
\m S=\frac{1}{(2-K^2(1-V^2))(1-2K^2(1-V^2))}
\!\pmatrix{N_{11}&N_{12}&N_{13}\cr N_{12}&N_{22}&N_{23}\cr N_{13}&N_{23}&N_{33}}\!,
\eeq
with
\beqa
N_{11}&=&2-K^2(1-V)(3+5V)+K^4(1-V)^2(1+2V^2),
\nonumber\\
N_{22}&=&(1+2K^2V^2)(2-K^2(1-V^2)),
\nonumber\\
N_{33}&=&2-K^2(1+V)(3-5V)+K^4(1+V)^2(1+2V^2),
\nonumber\\
N_{12}&=&K(2-K^2(1-V)(1+4V)),
\nonumber\\
N_{23}&=&K(2-K^2(1+V)(1-4V)),
\nonumber\\
N_{13}&=&K^2(2-K^2(1-V^2)).
\eeqa
The entropy production rate reads
\beq
\Phi=\frac{4K^2V^2(2+K^2V^2)}{2-K^2(1-V^2)}.
\label{phin3}
\eeq

\subsubsection*{$N=4$.}

The expression of the covariance matrix $\m S$ is too long to be reported here.
The entropy production rate reads
\beqa
\Phi&=&\frac{2K^2V^2}{(4-K^2(1-V^2))(4-5K^2(1-V^2))}
\nonumber\\
&\times&(48-40K^2+7K^4+2K^2(36+5K^2)V^2+15K^4V^4).
\label{phin4}
\eeqa

\subsection{Main features}
\label{cmain}

\begin{figure}[!ht]
\begin{center}
\includegraphics[angle=0,width=.6\linewidth]{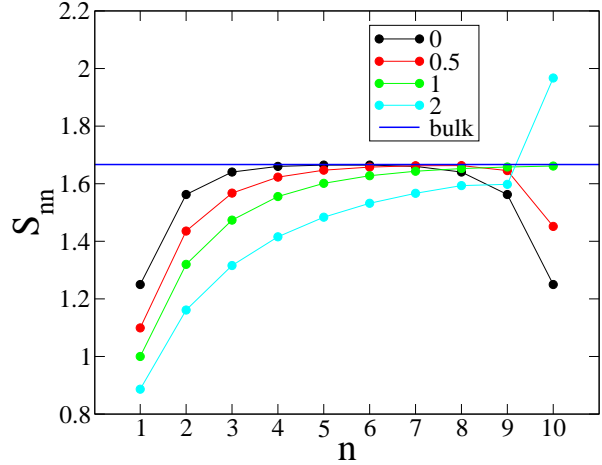}
\caption{\small
Plot of the spin strength $S_{nn}=\mean{x_n^2}$ against site number $n$
for an open chain of $N=10$ sites with $K=0.4$ and several $V$ (see legend).
The blue horizontal line shows the bulk spin strength~(\ref{smnbulk})
in the thermodynamic limit.}
\label{snn}
\end{center}
\end{figure}

The most salient novel features of the process on an open chain,
with respect to the geometry of a ring,
appear clearly on the above expressions for small values of $N$.
The nonequilibrium stationary state
now depends on the asymmetry parameter $V$,
and exhibits a non-trivial spatial dependence.
This is illustrated in figure~\ref{snn},
showing the spatial profile of the spin strength $S_{nn}=\mean{x_n^2}$
for an open chain of $N=10$ sites with $K=0.4$,
so that the correlation length is $1/\mu=1/\ln 2\approx1.4426$,
and several $V$.
In the case of a reversible process ($V=0$), we have $\m S=\m B^{-1}$.
The analytical result~(\ref{smnsym}) is shown in black.
As $V$ is increased,
the spin strength profile becomes more and more asymmetric,
and the trend to saturate to the thermodynamic value~(\ref{smnbulk})
is slower and slower.
For $V<1$, i.e., when $\m B$ has real spectrum,
the spin strength is less than its thermodynamic value
and decreases near the ends of the chain.
For $V>1$, i.e., when $\m B$ has complex spectrum,
the spin strength exhibits a pronounced overshoot near the right end.
Finally, for $V=1$, i.e., when $\m B$ is lower triangular,
the system does not `feel' its right end.
We shall come back to this totally asymmetric situation in section~\ref{vone}.

\begin{figure}[!ht]
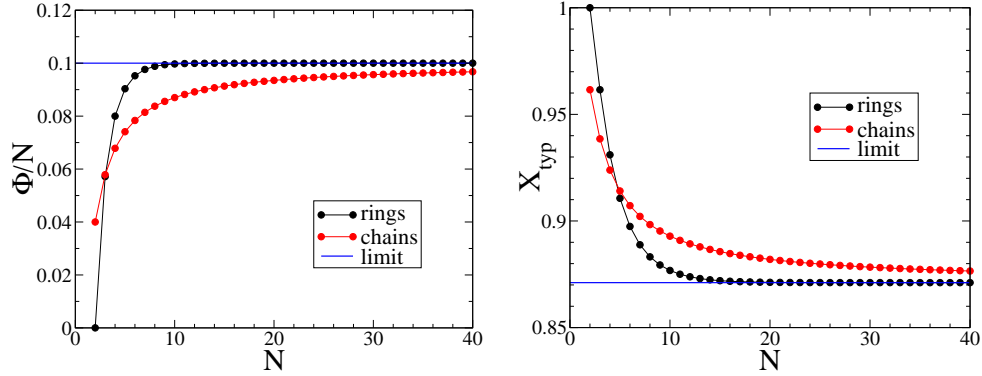

\begin{center}
\includegraphics[angle=0,width=.48\linewidth]{phicv.eps}
\hskip 10pt
\includegraphics[angle=0,width=.47\linewidth]{xcv.eps}
\caption{\small
Convergence of $\Phi/N$ (left) and $X\typ$ (right)
against $N$ on rings (black) and open chains (red)
for $K=0.4$ and $V=0.5$.
Blue horizontal lines show the thermodynamic limits~(\ref{philim}) and~(\ref{xlim}).}
\label{cv}
\end{center}
\end{figure}

In spite of the above,
intensive quantities characterising the nonequilibrium stationary state of the process
on an open chain and on a ring converge to the same thermodynamic limits.
This is illustrated in figure~\ref{cv},
showing the entropy production rate per spin $\Phi/N$ (left)
and the typical FDR $X\typ$ (right)
against system size $N$ up to 40 for rings (black) and open chains (red)
in a typical situation ($K=0.4$ and $V=0.5$).
The thermodynamic limits~(\ref{philim}) and~(\ref{xlim})
are shown as blue horizontal lines.
The convergence is observed to be very fast in the case of rings,
in agreement with the expected exponential convergence in $\e^{-N\mu}$.
In the geometry of an open chain,
extensive quantities are expected to possess a finite additive boundary contribution.
In the case of the entropy production rate, this reads
\beq
\Phi\approx N\varphi+\Phi\bdy,
\label{bandb}
\eeq
up to exponentially small corrections in $\e^{-N\mu}$.
Intensive quantities per spin are therefore expected to exhibit $1/N$ corrections.
This explains the observed slow convergence.

To close this study of the dynamics of the Gaussian model on an open chain,
we present more detailed
analytical investigations of the model in the small-$V$ regime
(section~\ref{smallv}) and in the totally asymmetric case $V=1$ (section~\ref{vone}).
The main emphasis will again be on the entropy production rate $\Phi$.

\subsection{The small-$V$ regime}
\label{smallv}

In the regime where the asymmetry parameter $V$ is small,
the entropy production rate vanishes quadratically.
This behavior appears clearly on the expressions~(\ref{phin2}),~(\ref{phin3})
and~(\ref{phin4}) for the first few values of $N$.

The entropy production amplitude
\beq
\Omega=\lim_{V\to0}\frac{\Phi}{V^2}
\label{odef}
\eeq
has been evaluated as a function of $K$ for all system sizes $N$.
The full derivation is given in~\ref{asmallv}.
We obtain (see~(\ref{aeres1}))
\beq
\Omega=\frac{16K^2}{(N+1)^2}\sum_{k+l\odd}
\frac{\sin^2q_k\sin^2q_l}{(1-K(\cos q_k+\cos q_l))(\cos q_k-\cos q_l)^2}.
\label{eres1}
\eeq

For large enough systems, the above result assumes the form (see~(\ref{bandb}))
\beq
\Omega\approx N\omega+\Omega\bdy,
\eeq
where the entropy production amplitude per spin (see~(\ref{aoser}))
\beq
\omega=1-\sqrt{1-4K^2}
\label{oser}
\eeq
agrees with~(\ref{philim}),
whereas the boundary contribution reads (see~(\ref{aomegabdy}))
\beq
\Omega\bdy=\omega+\frac{4}{\pi}({\bf E}(2K)-{\bf K}(2K)),
\label{omegabdy}
\eeq
where ${\bf E}$ and ${\bf K}$ are the complete elliptic integrals.
This boundary term is always negative.
It diverges logarithmically as
\beq
\Omega\bdy\approx\frac{4}{\pi}\left(\ln\frac{\mu}{4}+1\right)+1
\label{bdylog}
\eeq
as the critical point is approached ($\mu\to0$).

Right at the critical point, the entropy production amplitude scales as
(see~(\ref{acritlog}))
\beq
\Omega\approx N-\frac{4}{\pi}\ln\frac{N}{N_1}.
\label{critlog}
\eeq

It is very plausible that the logarithmic terms in~(\ref{bdylog}) and~(\ref{critlog})
are connected to each other by a finite-size scaling function
of the dimensionless variable $N\mu$.
As a first step in this direction,
let us mention -- without giving any detail --
that we have been able to evaluate the finite part $N_1$
entering~(\ref{critlog}) and found
\beq
N_1=\frac{\Gamma^2(1/4)}{16\sqrt{\pi}}\,\exp\!\left(\frac{\pi}{4}+\frac{1}{2}-\euler\right)
\approx0.941080,
\label{n1}
\eeq
where $\euler$ is Euler's constant.

\subsection{The totally asymmetric case $V=1$}
\label{vone}

In the borderline case $V=1$, where interactions are totally asymmetric,
the friction matrix $\m B$ is lower triangular.
The Sylvester equation~(\ref{sexpl}) simplifies to the recursion
\beq
S_{mn}=K(S_{m-1,n}+S_{m,n-1})+\delta_{mn},
\label{srec}
\eeq
whose solution reads
\beq
S_{mn}=\sum_{k=1}^{\min(m,n)}\frac{(m+n-2k)!}{(m-k)!(n-k)!}\,K^{m+n-2k}.
\label{smn}
\eeq
This expression is readily checked by inspection.
It does not involve the system size~$N$ explicitly.
Indeed, as already observed on figure~\ref{snn},
the system does not `feel' its right end.

Far from the left boundary, the translationally invariant correlations
of the infinite system should be recovered.
For $m$ and $n$ large, keeping the distance $l=n-m\ge0$ fixed,
and setting $j=m-k$, we indeed find that the expression~(\ref{smn}) converges to
\beq
S_l=\sum_{j\ge0}\frac{(2j+l)!}{j!(j+l)!}\,K^{2j+l}=\cotanh\mu\,\e^{-\mu l},
\eeq
in agreement with~(\ref{snbulk}),~(\ref{smnbulk}).

The entries of the antisymmetric matrix $\m Q$ read
\beq
Q_{mn}=K(S_{m,n-1}-S_{m-1,n}),
\eeq
hence
\beq
Q_{mn}=(n-m)\sum_{k=1}^{\min(m,n)}\frac{(m+n-2k-1)!}{(m-k)!(n-k)!}\,K^{m+n-2k}.
\label{qmn}
\eeq
Far from the left boundary, with the above conventions,
we have $Q_l=-Q_{-l}=\e^{-\mu l}$ for $l\ge1$, and $Q_0=0$.

The entropy production rate reads (see~(\ref{phi2}))
\beq
\Phi=2K\sum_{n=1}^{N-1}Q_{n,n+1},
\eeq
where the dependence on the size $N$ is entirely contained in the summation bound.
Using~(\ref{qmn}) and rearranging the sums, we obtain
\beq
\Phi=2\sum_{j=0}^{N-1}(N-1-j)\frac{(2j)!}{j!(j+1)!}\,K^{2j+2}.
\label{phiv1}
\eeq
This result can be readily brought to the form~(\ref{bandb}) for large chains.
For the entropy production rate per spin $\varphi$,
we obtain the very same series expansion~(\ref{oser}) as in the small-$V$ regime.
The expression~(\ref{philim}) is thus recovered.
For the boundary contribution, we obtain
\beq
\Phi\bdy=-2\sum_{j\ge0}\frac{(2j)!}{j!^2}\,K^{2j+2}
=-\frac{2K^2}{\sqrt{1-4K^2}}=-\frac{1}{\sinh 2\mu}.
\label{phiv1bdy}
\eeq
This boundary term is always negative.
It diverges linearly as the critical point is approached ($\mu\to0$),
whereas its counterpart $\Omega\bdy$ in the small-$V$ regime
diverges only logarithmically (see~(\ref{bdylog})).

It is worth investigating the totally asymmetric case right at the critical point.
The spin strength at site $n$ diverges with distance $n$ according~to
\beq
S_{nn}=\sum_{j=0}^{n-1}\frac{(2j)!}{2^{2j}j!^2}\approx2\sqrt\frac{n}{\pi}.
\eeq
This square-root divergence is however weaker than the linear growth of the spin strength
in the bulk of the chain at criticality in the symmetric case (see~(\ref{ccrit})), i.e.,
\beq
S_{nn}=\frac{2n(N+1-n)}{N+1}.
\eeq
The expression~(\ref{phiv1}) for the entropy production rate can be recast as
\beq
\Phi=N-\frac{1}{2}\sum_{j=0}^{N-1}\frac{(2j)!}{2^{2j}j!^2}
-\frac{N}{2}\sum_{j=N}^\infty\frac{(2j)!}{2^{2j}j!(j+1)!}.
\eeq
Each correction term behaves asymptotically as $\sqrt{N/\pi}$, and so
\beq
\Phi\approx N-2\sqrt\frac{N}{\pi}.
\label{phiv1c}
\eeq

Finally, the finite-size scaling law obeyed by the entropy production rate
in the critical regime ($N$ large and $\mu$ small)
can also be evaluated.
Simplifying the expression~(\ref{phiv1}) for $\mu$ small,
replacing sums by integrals, we indeed obtain
\beq
\Phi\approx N-\sqrt{N}\,F(x),
\eeq
where, rather unexpectedly, the argument of the scaling function is $x=\sqrt{N}\mu$,
and not $N\mu$, which is the natural scaling variable.
The scaling function is given by
\beq
F(x)=(x^2+1)\frac{\erf x}{x}+\frac{\e^{-x^2}}{\sqrt{\pi}}.
\eeq
For $x\ll1$, the expansion
\beq
F(x)=\frac{2}{\sqrt{\pi}}\left(1+\frac{x^2}{3}+\cdots\right)
\eeq
reproduces the critical behavior~(\ref{phiv1c}), with regular corrections.
For $x\gg1$, we have
\beq
F(x)\approx x+\frac{1}{2x}-\frac{2\e^{-x^2}}{\sqrt{\pi}}.
\eeq
The first term matches the leading behavior
of the thermodynamic result~(\ref{philim}), i.e., $\varphi\approx(1-\mu)V^2$.
The second term matches the divergence of the expression~(\ref{phiv1bdy})
of~$\Phi\bdy$ as $\mu\to0$.
The third term shows that the leading correction
to the asymptotic result~(\ref{bandb}) is indeed exponentially small.

\section{Electrical arrays}
\label{elec}

Our second class of examples of multivariate Ornstein-Uhlenbeck processes
originates in macroscopic physics.
We shall consider arrays of $N$ resistively coupled
$RL$ and $RC$ electrical circuits,
where resistors are sources of thermal noise.
Within this framework, irreversibility is brought about by an inhomogeneous
temperature profile~\cite{LTO,BMN}.
Electrical circuits have been used in the past as a playground to test
general ideas around the minimum and maximum entropy production
principles~\cite{L1,L2}.

\subsection{$RL$ network}
\label{rl}

\begin{figure}[!ht]
\begin{center}
\includegraphics[angle=0,width=.6\linewidth]{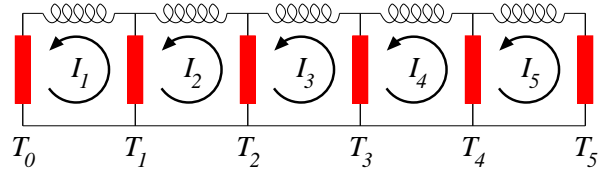}
\caption{\small
An array of $N=5$ resistively coupled $RL$ electrical circuits.}
\label{rlfig}
\end{center}
\end{figure}

We first consider an array of $N$ resistively coupled $RL$ circuits,
as depicted in figure~\ref{rlfig}.
All resistors and coils have the same resistance $R$ and inductance $L$.
The loop currents are denoted as $I_n(t)$ ($n=1,\dots,N$).
In the most general situation,
each resistor is kept at a different temperature $T_n$ ($n=0,\dots,N$).

The total voltage drop around the loop traversed by the current $I_n(t)$ reads
\beqa
\Delta V_n(t)&=&R(I_n(t)-I_{n+1}(t))+\xi_n(t)+L\frac{\dd I_n(t)}{\dd t}
\nonumber\\
&-&R(I_{n-1}(t)-I_n(t))-\xi_{n-1}(t)=0.
\label{deltavrl}
\eeqa
According to the Johnson-Nyquist theory of thermal noise,
the $\xi_n(t)$ are independent Gaussian white noises obeying
\beq
\mean{\xi_m(t)\xi_n(t')}=2RT_n\delta_{mn}\delta(t-t').
\label{xidef}
\eeq
The Dirichlet boundary conditions $I_0(t)=I_{N+1}(t)=0$ hold throughout section~\ref{elec}.

Equation~(\ref{deltavrl}) can be recast as
\beq
\frac{\dd I_n(t)}{\dd t}=-\frac{R}{L}(2I_n(t)-I_{n-1}(t)-I_{n+1}(t))+\eta_n(t),
\label{didt}
\eeq
with
\beq
\eta_n(t)=\frac{1}{L}\left(\xi_{n-1}(t)-\xi_n(t)\right).
\label{etadef}
\eeq
The above equations define an Ornstein-Uhlenbeck process of the form~(\ref{ou})
in terms of the intensities $I_n(t)$.
The friction matrix reads
\beq
\m B=\frac{R}{L}\,\m\Delta,
\eeq
where $\m\Delta$ is (minus) the Laplacian on the chain.
It is a symmetric tridiagonal Toeplitz matrix whose non-zero elements are
\beqa
\Delta_{nn}=2,
\nonumber\\
\Delta_{n,n-1}=\Delta_{n-1,n}=-1\qquad(n\ne1).
\label{deltaelts}
\eeqa
We remark that $\m\Delta$
is twice the matrix $\m A$ giving the action of the critical Gaussian spin model
on the open chain.
As a consequence, the entries of its inverse read (see~(\ref{ccrit}))
\beq
(\m\Delta^{-1})_{mn}=\frac{m(N+1-n)}{N+1}\qquad(m\le n).
\label{lin}
\eeq
We have in particular
\beq
\tr\m\Delta^{-1}=\frac{N(N+2)}{6}.
\label{sumrule}
\eeq

The diffusion matrix reads
\beq
\m D=\frac{R}{L^2}\m\T,
\eeq
where $\m\T$ is the temperature matrix,
whose non-zero elements can be worked out from~(\ref{xidef}) and~(\ref{etadef}):
\beqa
\T_{nn}=T_{n-1}+T_n,
\nonumber\\
\T_{n,n-1}=\T_{n-1,n}=-T_{n-1}\qquad(n\ne1).
\label{telts}
\eeqa
As a consequence of the above, the Sylvester equation~(\ref{sid}) reads
\beq
\m\Delta\m S+\m S\m\Delta=\frac{2}{L}\,\m\T.
\label{sidrl}
\eeq

When all resistors are at the same temperature ($T_n=T$), we have
\beq
\m\T=T\m\Delta,
\label{thomo}
\eeq
the matrices $\m B$ and $\m D$ are proportional to each other,
and the process is reversible, as expected.
The covariance matrix reads
\beq
\m S=\frac{T}{L}\,\m 1,
\eeq
and so the loop intensities $I_n(t)$ are independent and equally distributed
Gaussian processes.

As recalled above, irreversibility is brought about by the spatial inhomogeneity
of the temperature profile.
Indeed, as soon as temperature is not homogeneous,
the matrices $\m B$ and $\m D$ do not commute,
the process is irreversible, and the Sylvester equation~(\ref{sidrl})
for the covariance matrix $\m S$ is non-trivial.
The case $N=2$ is however an exception:
$\m B$ and~$\m D$ commute and the process is reversible,
even in the presence of an inhomogeneous symmetric temperature profile,
such that $T_0=T_2\ne T_1$ (see below).

The main emphasis will again be on the entropy production rate $\Phi$.
Furthermore,
we are mostly interested in the regime where temperature fluctuations are small,
i.e.,
\beq
T_n=\Tbar+\delta T_n\qquad(\abs{\delta T_n}\ll\Tbar).
\label{dtndef}
\eeq
In this regime, let us anticipate that
the entropy production rate is given by a quadratic form
of the temperature fluctuations, i.e.,
\beq
\Phi\approx\frac{R}{L\Tbar^2}\sum_{m,n=0}^N\Omega_{mn}\delta T_m\delta T_n.
\label{ophidef}
\eeq
For the sake of clarity, the bounds of summations
will be written down explicitly throughout this section~\ref{elec}.
The amplitude matrix $\m\Omega$ is dimensionless and only depends on the network size $N$.
It is a positive semi-definite matrix
with dimension $(N+1)\times(N+1)$ and rank $N$ (rank~1 for $N=2$).

In the special situation of a weakly disordered temperature profile,
where the temperature fluctuations $\delta T_n$
are independent and identically distributed random variables such that
\beq
\mean{\delta T_n}=0,\qquad
\mean{\delta T_m\delta T_n}=w^2\delta_{mn},
\label{disdef}
\eeq
with $w\ll\Tbar$, the result~(\ref{ophidef}) yields the following prediction
for the mean entropy production rate:
\beq
\mean{\Phi}\approx\frac{Rw^2}{L\Tbar^2}\,\Lambda,
\label{phidis}
\eeq
with
\beq
\Lambda=\tr\m\Omega.
\eeq

It is worth having a closer look at the first few values of $N$.

\subsubsection*{$N=2$.}

We have
\beq
\m B=\frac{R}{L}\pmatrix{2&-1\cr -1&2},\qquad
\m D=\frac{R}{L^2}\pmatrix{T_0+T_1&-T_1\cr -T_1&T_1+T_2}.
\eeq
The reversibility condition $\m B\m D=\m D\m B$ yields one single condition: $T_0=T_2$.
As a consequence, $\m B$ and $\m D$ commute
and the process is reversible,
even in the presence of an inhomogeneous symmetric temperature profile,
such that $T_0=T_2\ne T_1$.
The Sylvester equation~(\ref{sidrl}) yields
\beq
\m S=\frac{1}{12L}\pmatrix{7T_0+4T_1+T_2&2(T_0-2T_1+T_2)\cr 2(T_0-2T_1+T_2)&T_0+4T_1+7T_2}.
\label{srl2}
\eeq
We have then
\beq
\m Q=\frac{R(T_2-T_0)}{4L^2}\pmatrix{0&-1\cr 1&0}.
\eeq
The entropy production rate reads, in full generality
\beq
\Phi=\frac{R(T_2-T_0)^2}{4L(T_0T_1+T_0T_2+T_1T_2)}.
\label{phirl2}
\eeq
When the temperature difference $T_2-T_0$ is small,
this expression simplifies to
\beq
\Phi\approx\frac{R(T_2-T_0)^2}{12L\Tbar^2},
\eeq
in agreement with the announced form~(\ref{ophidef}), with
\beq
\m\Omega=\frac{1}{12}\pmatrix{1&0&-1\cr 0&0&0\cr -1&0&1},
\eeq
and therefore
\beq
\Lambda=\frac{1}{6}.
\eeq

\subsubsection*{$N=3$.}

We have
\beqa
\m B&=&\frac{R}{L}\pmatrix{2&-1&0\cr -1&2&-1\cr 0&-1&2},
\nonumber\\
\m D&=&\frac{R}{L^2}\pmatrix{T_0+T_1&-T_1&0\cr -T_1&T_1+T_2&-T_2\cr 0&-T_2&T_2+T_3}.
\eeqa
The general expressions for $\m S$ and $\Phi$ are too lengthy to be reported here.
The amplitude matrix
\beq
\m\Omega=\frac{1}{112}\pmatrix{17&-9&-5&-3\cr -9&13&1&-5\cr -5&1&13&-9\cr -3&-5&-9&17}
\eeq
has rank 3 and trace
\beq
\Lambda=\frac{15}{28}.
\eeq

\subsubsection*{$N=4$.}

We shall only give the expressions of the amplitude matrix
\beq
\m\Omega=\frac{1}{3300}\pmatrix{653&-437&-132&-47&-37\cr
-437&698&-132&-82&-47\cr -132&-132&528&-132&-132\cr
-47&-82&-132&698&-437\cr -37&-47&-132&-437&653}
\eeq
and of its trace
\beq
\Lambda=\frac{323}{330}.
\eeq

For an arbitrary network size $N$,
the amplitude matrix $\m\Omega$ entering the expres\-sion~(\ref{ophidef})
of the entropy production rate
has been derived in~\ref{arl}.
The outcome reads (see~(\ref{aomegares}))
\beqa
\Omega_{mn}&=&\frac{4}{(N+1)^2}
\sum_{kl}\frac{(\cos q_k-\cos q_l)^2}{2-\cos q_k-\cos q_l}
\nonumber\\
&\times&\cos(m+\half)q_k\cos(m+\half)q_l\cos(n+\half)q_k\cos(n+\half)q_l.
\label{omegares}
\eeqa

The quantity $\Lambda=\tr\m\Omega$ can be worked out in more detail.
We thus obtain
\beq
\Lambda=-\frac{N-1}{2(N+1)}+\frac{1}{N+1}
\sum_{kl}\frac{(\cos q_k-\cos q_l)^2}{2-\cos q_k-\cos q_l}.
\label{lambdares}
\eeq
For a large network ($N\to\infty$), we have (see~(\ref{bandb}))
\beq
\Lambda\approx N\lambda+\Lambda\bdy.
\label{lbandb}
\eeq
The intensive part $\lambda$ is readily obtained by replacing
the double sum in~(\ref{lambdares}) by the corresponding integral.
We thus get
\beq
\lambda=\int_0^\pi\frac{\dd x}{\pi}\int_0^\pi\frac{\dd y}{\pi}
\frac{(\cos x-\cos y)^2}{2-\cos x-\cos y}
=\frac{8}{\pi}-2\approx0.546479.
\eeq
The boundary contribution $\Lambda\bdy$ can be derived by means of a careful analysis
of the difference between the double sum in~(\ref{lambdares})
and the corresponding integral, using twice the first-order Euler-Maclaurin formula.
We are left with
\beq
\Lambda\bdy=\frac{8}{\pi}+\frac{3}{2}-4\sqrt{2}\approx-1.610375.
\eeq
This quantity is negative, just as its counterparts~(\ref{omegabdy})
and~(\ref{phiv1bdy}) in the Gaussian spin model on an open chain.
In all these cases,
there is less entropy production near the ends of the sample than in its bulk.

To close, let us consider the situation
where the temperature profile is homo\-geneous over the whole network ($T_n=\Tbar$),
except for the leftmost resistor,
whose temperature is $T_0=\Tbar+\delta T_0$, with $\abs{\delta T_0}\ll\Tbar$.
We have then
\beq
\Phi\approx\frac{R(\delta T_0)^2}{L\Tbar^2}\,\Omega_{00}.
\eeq
For a large enough network ($N\to\infty$),
the expression~(\ref{omegares}) for the amplitude~$\Omega_{00}$ approaches the limit
\beqa
\Omega_{00}&=&\int_0^\pi\frac{\dd x}{\pi}\int_0^\pi\frac{\dd y}{\pi}
\frac{(\cos x-\cos y)^2(1+\cos x)(1+\cos y)}{2-\cos x-\cos y}
\nonumber\\
&=&\frac{32}{3\pi}-3\approx0.395305.
\label{w00}
\eeqa
The inequality $\Omega_{00}<\lambda$ confirms that the local entropy production rate
is smaller near the boundaries.

\subsection{$RC$ network}
\label{rc}

\begin{figure}[!ht]
\begin{center}
\includegraphics[angle=0,width=.6\linewidth]{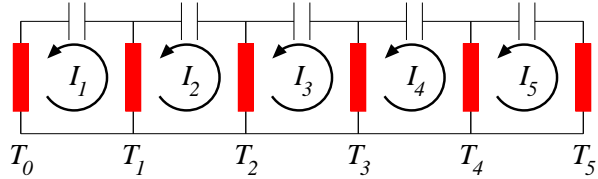}
\caption{\small
An array of $N=5$ resistively coupled $RC$ electrical circuits.}
\label{rcfig}
\end{center}
\end{figure}

We now consider an array of $N$ resistively coupled $RC$ circuits,
as depicted in figure~\ref{rcfig}.
All resistors and capacitors have the same values $R$ and $C$.
The loop currents read
\beq
I_n(t)=\frac{\dd Q_n(t)}{\dd t}\qquad(n=1,\dots,N),
\eeq
where $Q_n(t)$ is the electric charge of the right electrode of the capacitor
of circuit number $n$.
As in the previous case,
each resistor is kept at a different temperature $T_n$ ($n=0,\dots,N$).

The total voltage drop around the loop traversed by $I_n(t)$ reads
\beqa
\Delta V_n(t)&=&R(I_n(t)-I_{n+1}(t))+\xi_n(t)+\frac{Q_n(t)}{C}
\nonumber\\
&-&R(I_{n-1}(t)-I_n(t))-\xi_{n-1}(t)=0,
\label{deltavrc}
\eeqa
where the Johnson-Nyquist noises $\xi_n(t)$ are still given by~(\ref{xidef}).
The above equations can be recast as
\beq
Q_n(t)=-RC\,\frac{\dd}{\dd t}\left(2Q_n(t)-Q_{n-1}(t)-Q_{n+1}(t)\right)+\alpha_n(t),
\label{dqdt}
\eeq
with
\beq
\alpha_n(t)=C\left(\xi_{n-1}(t)-\xi_n(t)\right)=LC\,\eta_n(t).
\label{alphadef}
\eeq
The similarity between equations~(\ref{didt}) and~(\ref{dqdt})
demonstrates that the $RL$ and $RC$ networks are somehow dual to each other.
The noise terms are proportional to each other,
whereas time derivatives act on the left-hand side in~(\ref{didt}),
and on the right-hand side in~(\ref{dqdt}).

Equations~(\ref{dqdt}) can be brought to the canonical form~(\ref{ou})
of a multivariate Ornstein-Uhlenbeck process in terms of the charges $Q_n(t)$,
by acting upon them by the inverse of the Laplacian $\m\Delta$.
We thus obtain
\beq
\frac{\dd Q_m(t)}{\dd t}=-\frac{1}{RC}\sum_{n=1}^N
(\m\Delta^{-1})_{mn}\left(Q_n(t)-\alpha_n(t)\right).
\eeq
At variance with the examples considered so far in this work,
this process is non-local,
in the sense that the corresponding matrices, i.e.,
\beq
\m B=\frac{1}{RC}\,\m\Delta^{-1},\qquad
\m D=\frac{1}{R}\,\m\Delta^{-1}\m\T\m\Delta^{-1},
\eeq
are long-ranged.
The temperature matrix $\m\T$ is again given by~(\ref{telts}).
As a consequence of the above, the Sylvester equation~(\ref{sid}) reads
\beq
\m\Delta\m S+\m S\m\Delta=2C\m\T.
\label{sidrc}
\eeq
The similarity between equations~(\ref{sidrl}) and~(\ref{sidrc})
yields a remarkable consequence of the above duality.
If the $RL$ and $RC$ networks are subjected to the same temperature profile,
their covariance matrices are proportional to each other, i.e.,
\beq
\m S_{RC}=LC\,\m S_{RL},
\label{sdual}
\eeq
with obvious notations.
The above symmetry however does not extend to all obser\-vables.
In particular, there is no simple relationship between the entropy production rates
of the two networks.

When all resistors are at the same temperature ($T_n=T$), we have
$\m\T=T\m\Delta$ (see~(\ref{thomo})), hence
\beq
\m D=\frac{T}{R}\,\m\Delta^{-1}.
\eeq
Here again,
the matrices $\m B$ and $\m D$ are proportional to each other,
and the process is reversible, as expected.
The covariance matrix reads
\beq
\m S=CT\,\m 1,
\eeq
and so the charges $Q_n(t)$ are independent and equally distributed Gaussian processes.

The main quantities of interest will again be the entropy production rate $\Phi$
in the regime where temperature fluctuations are small (see~(\ref{dtndef})),
which now reads
\beq
\Phi\approx\frac{1}{RC\Tbar^2}\sum_{m,n=0}^N\Omega_{mn}\delta T_m\delta T_n,
\label{rcodef}
\eeq
where the amplitude matrix $\m\Omega$ is again a dimensionless
and positive semi-definite matrix, which only depends on size $N$.
In the special situation of a weakly disordered temperature profile (see~(\ref{disdef})),
this becomes
\beq
\mean{\Phi}\approx\frac{w^2}{RC\Tbar^2}\,\Lambda,
\label{phidisrl}
\eeq
where again
\beq
\Lambda=\tr\m\Omega.
\eeq

Before we proceed, it is again worth looking at the first few values of $N$.

\subsubsection*{$N=2$.}

The expression~(\ref{lin}) yields
\beqa
\m B&=&\frac{1}{3RC}\pmatrix{2&1\cr 1&2},
\nonumber\\
\m D&=&\frac{1}{9R}\pmatrix{4T_0+T_1+T_2&2T_0-T_1+2T_2\cr 2T_0-T_1+2T_2&T_0+T_1+4T_2}.
\eeqa
The reversibility condition $\m B\m D=\m D\m B$ again yields the single condition $T_0=T_2$.
The matrix $\m S$ is proportional to~(\ref{srl2}), in agreement with~(\ref{sdual}).
We have then
\beq
\m Q=\frac{T_2-T_0}{12R}\pmatrix{0&-1\cr 1&0}.
\eeq
The entropy production rate reads
\beq
\Phi=\frac{(T_2-T_0)^2}{12RC(T_0T_1+T_0T_2+T_1T_2)}.
\eeq
When the temperature difference $T_2-T_0$ is small,
$\Phi$ has the form~(\ref{rcodef}), with
\beq
\m\Omega=\frac{1}{36}\pmatrix{1&0&-1\cr 0&0&0\cr -1&0&1},
\eeq
and therefore
\beq
\Lambda=\frac{1}{18}.
\eeq

\subsubsection*{$N=3$.}

We shall only give the expression of the amplitude matrix
\beq
\m\Omega=\frac{1}{224}\pmatrix{23&-13&-1&-9\cr -13&11&3&-1\cr -1&3&11&-13\cr -9&-1&-13&23}
\eeq
and of its trace
\beq
\Lambda=\frac{17}{56}.
\eeq

\subsubsection*{$N=4$.}

We shall only give the expression of the amplitude matrix
\beq
\m\Omega=\frac{1}{16500}\pmatrix{3583&-2392&-132&-292&-767\cr
-2392&2683&-132&133&-292\cr -132&-132&528&-132&-132\cr
-292&133&-132&2683&-2392\cr -767&-292&-132&-2392&3583}
\eeq
and of its trace
\beq
\Lambda=\frac{653}{825}.
\eeq

For an arbitrary network size $N$,
the amplitude matrix $\m\Omega$ entering the expres\-sion~(\ref{rcodef})
of the entropy production rate
has been derived in~\ref{arc}.
The outcome reads (see~(\ref{arcomegares}))
\beqa
\Omega_{mn}&=&\frac{1}{(N+1)^2}
\sum_{kl}\frac{(\cos q_k-\cos q_l)^2}{(1-\cos q_k)(1-\cos q_l)(2-\cos q_k-\cos q_l)}
\nonumber\\
&\times&\cos(m+\half)q_k\cos(m+\half)q_l\cos(n+\half)q_k\cos(n+\half)q_l.
\label{rcomegares}
\eeqa

The quantity $\Lambda$ can be worked out in more detail.
We thus obtain
\beq
\Lambda=\frac{1}{4(N+1)}
\sum_{kl}\frac
{(1-\half\delta_{k+l,N+1})(\cos q_k-\cos q_l)^2}
{(1-\cos q_k)(1-\cos q_l)(2-\cos q_k-\cos q_l)}.
\eeq
Some algebra using the sum rule~(\ref{sumrule}) yields
\beq
\Lambda=\frac{N(2N+1)}{12}-\frac{C_N}{N+1},
\eeq
with
\beq
C_N=\sum_{kl}\frac{1}{2-\cos q_k-\cos q_l}.
\eeq
For a large network ($N\to\infty$),
expanding the cosines to quadratic order,
the latter sum can be estimated as
\beq
C_N\approx\frac{2N^2}{\pi^2}\sum_{kl}\frac{1}{k^2+l^2}
\sim\frac{N^2}{\pi}\ln N.
\eeq
We thus obtain
\beq
\Lambda\approx\frac{N^2}{6}-\frac{N}{\pi}\ln\frac{N}{N_2}.
\label{lambdacrit}
\eeq
The finite part $N_2$ is not predicted by this heuristic reasoning.

In the situation
where the temperature profile is homogeneous except for the leftmost resistor,
whose temperature is $T_0=\Tbar+\delta T_0$, with $\abs{\delta T_0}\ll\Tbar$,
we have
\beq
\Phi\approx\frac{(\delta T_0)^2}{RC\Tbar^2}\,\Omega_{00}.
\eeq
For a large network ($N\to\infty$),
the expression~(\ref{rcomegares}) for~$\Omega_{00}$
can be analysed along the previous lines.
We thus obtain
\beq
\Omega_{00}\approx\frac{N}{3}-\frac{4}{\pi}\ln\frac{N}{N_3},
\qquad N_3=N_2\exp\left(\frac{\pi}{24}+\frac{1}{2}\right).
\label{omegacrit}
\eeq

The entropy production rate exhibits an anomalous growth with the network size~$N$.
In the case of a weakly disordered temperature profile,
the amplitude $\Lambda$ grows quadratically (see~(\ref{lambdacrit})),
whereas it grows linearly in the $RL$ network (see~(\ref{lbandb})).
In the case of a boundary temperature inhomogeneity,
the amplitude $\Omega_{00}$ grows linearly (see~(\ref{omegacrit})),
whereas it saturates to a constant value in the $RL$ network (see~(\ref{w00})).
These anomalous growth laws reflect a violation of extensivity
which can be attributed to the peculiar feature of the present $RC$ network,
namely that the process is non-local,
in the sense that the matrices $\m B$ and $\m D$ are long-ranged.
As far as logarithmic subleading terms are concerned,
they are due to the effective criticality of the model.
Indeed, logarithmic corrections are also met
in the $V\to0$ regime of the critical open spin chain
(see~(\ref{bdylog}) and~(\ref{critlog})),
but not in the case of the $RL$ network.

To close, let us mention -- again without giving any detail --
that we have evaluated the finite part $N_2$ and found
\beq
N_2=\frac{\Gamma^2(1/4)}{8\sqrt{2\pi}}\,\exp\!\left(\frac{5\pi}{12}-\euler\right)
\approx1.362670,
\eeq
where $\euler$ is again Euler's constant.
Note the striking similarity with~(\ref{n1}).

\section{Discussion}
\label{discussion}

This work was devoted to a characterisation
of the nonequilibrium stationary state
of a generic multivariate Ornstein-Uhlenbeck process involving $N$ degrees of freedom.
The main advantage of this class of processes
is that the linearity of the Langevin equations allows us to use linear algebra,
and therefore to derive closed-form expressions for many quantities of interest.

The key point unifying the general results derived in section~\ref{general}
is that the irreversibility of the process is characterised by
a single antisymmetric matrix~$\m Q$, which is nothing but the antisymmetric part
of the Onsager matrix $\m L$ of kinetic coefficients (see~(\ref{qdef})).
All physical quantities
characterising the nonequilibrium stationary state have been expressed either
in terms of the matrix $\m Q$ or of its associated Hermitian form~$\m H$ (see~(\ref{hdef})).
These quantities include, most importantly,
the entropy production rate $\Phi$ (see~(\ref{phi1}),~(\ref{phi2})),
as well as the stationary probability current~$\m J(\m x)$ (see~(\ref{eq:J}),~(\ref{mudef})),
the fluctuation-dissipation ratio matrix~$\m X$ (see~(\ref{xres}))
and typical fluctuation-dissipation ratio $X\typ$ (see~(\ref{xtyp})),
and the asymmetry index $\w H\imax$ (see~(\ref{hmax})).

Our first example of multivariate Ornstein-Uhlenbeck processes
is the one-dimensional ferromagnetic Gaussian spin model
endowed with a stochastic dynamics where spatial asymmetry results in irreversibility.
This dynamics is the most general one
keeping both the linearity and the range of interactions.
Its parametrisation is inspired by earlier studies
on directed Ising and spherical ferromagnets~\cite{gb2009,cg2011,cg2013,GLI,GLSA}.
We successively investigated the model
on a ring (section~\ref{ring}) and on an open chain (section~\ref{chain}).
There is a qualitative difference between the two geometries.
The stationary state of the model on a ring is independent of the asymmetry parameter~$V$,
whereas it depends continuously on $V$ on an open chain.
The independence of stationary-state spin correlations on $V$ in the ring geometry
can be attributed to a symmetry: the process is cyclically symmetric.
The very same property had been observed on the spherical model
in the thermodynamic limit in any dimension~\cite{GLSA}.
There too, the independence of stationary correlations on the bias
originates in translational invariance.
The influence of symmetries on the stationary state of Gaussian and spherical models
is by no means limited to the realm of stochastic dynamics.
Let us mention a static analogue of the above.
The well-known equivalence between the O$(N)$ Heisenberg model
in the large-$N$ limit and the spherical model~\cite{stanley}
only holds if the sample is a symmetric space, i.e., if all sites are equivalent.
In other geometries, such as the open chain considered in the present work,
the situation is more complex.
The Heisenberg model indeed identifies, in the large-$N$ limit,
with a generalised spherical model
involving as many constraints as there are classes of inequivalent sites~\cite{knops}.

The general results obtained in section~\ref{general}
allowed us to derive explicit expressions for the physical quantities
characterising the nonequilibrium stationary state of the Gaussian spin model
in the thermodynamic limit.
All these results (see section~\ref{rthermo}) are independent of the geometry.
Much attention has also been paid to finite-size effects,
which are different in the ring and chain geometries,
especially in the vicinity of the critical point.
The main focus has been on the entropy production rate~$\Phi$,
which is both fundamental
for the characterisation of a nonequilibrium stationary state,
and the easiest to analyse.
In the critical region on a ring,
$\Phi$ is the sum of an extensive part
and of a finite-size contribution which is a function of the natural scaling variable~$N\mu$,
where $\mu$ is the inverse of the static correlation length.
The situation on an open chain is more contrasted.
First of all,~$\Phi$ is the sum of an extensive part
and of a finite, negative boundary contribution (see~(\ref{bandb})).
The latter boundary term diverges as the critical point is approached,
albeit in a non-universal fashion,
namely logarithmically (either in $N$ or in $\mu$)
in the regime of a small asymmetry ($V\to0$),
and as a square-root of $N$ in the case of totally asymmetric interactions ($V=1$).

For our second class of examples of multivariate Ornstein-Uhlenbeck processes,
we have taken our inspiration from macroscopic physics.
We have considered arrays of $N$ resistively coupled $RL$ and $RC$ electrical circuits.
The main emphasis has been put on the entropy production rate
in the situation where the local temperatures of the resistors
are independent random variables with small fluctuations.
As a general rule,
there is less entropy production near the ends of the networks than in their bulk.
Another salient outcome is the qualitatively different scaling
of the entropy production rate in both cases.
On $RL$ networks (section~\ref{rl}),
$\Phi$ is again the sum of an extensive part
and of a negative boundary contribution (see~(\ref{lbandb})).
On $RC$ networks (section~\ref{rc}),
$\Phi$ grows quadratically with $N$, thus violating extensivity,
with a large negative `non-local boundary' contribution, growing as $N\ln N$
(see~(\ref{lambdacrit})).

Irreversible stochastic processes manifest yet another generic feature.
Asymmetry is known to accelerate the dynamics
of diffusions and other classes of processes~\cite{gb2009,cg2011,gp1,gp2,gp3,HHS1,HHS2,LN,WH,jack}.
This phenomenon is however absent in most of the examples considered in this work.
For the Gaussian spin model on a ring, this is because the process is cyclically symmetric.
For the electrical networks, this is because of the symmetry of Kirchhoff's laws.
The above acceleration is therefore only observed
in the ferromagnetic Gaussian model on an open chain (see~(\ref{gapchain})).
We hope to come back to this acceleration mechanism in a more general setting in future work.

\appendix

\section{Derivation of the expression~(\ref{oser})
for the entropy production amplitude on the open spin chain}
\label{asmallv}

In this appendix we give a full derivation of the expression~(\ref{oser})
for the entropy production amplitude $\Omega$
of the spin model on an open chain in the regime where the asymmetry parameter $V$ is small.
There, it is natural to expand quantities of interest as power series in $V$.
Setting (see~(\ref{belts}))
\beq
\m B=\m B\ze+\m B\un V,
\eeq
we look for a solution to the Sylvester equation~(\ref{sid}) in the form
\beq
\m S=\m S\ze+\m S\un V+\cdots
\eeq
We have $\m S\ze=(\m B\ze)^{-1}$, whose entries are given by~(\ref{smnsym}),
whereas $\m S\un$ obeys
\beq
\m B\ze\m S\un+\m S\un\m B\ze=\m S\ze\m B\un-\m B\un\m S\ze.
\label{sone}
\eeq
We have then $\m Q=\m Q\un V+\cdots$, where
\beq
\m Q\un=\m B\un\m S\ze+\m B\ze\m S\un=\m S\ze\m B\un-\m S\un\m B\ze.
\label{q1}
\eeq
The entropy production amplitude (see~(\ref{odef}))
\beq
\Omega=\lim_{V\to0}\frac{\Phi}{V^2}
\eeq
is given by (see~(\ref{phi2}))
\beq
\Omega=-\tr(\m B\un\m Q\un).
\label{eres}
\eeq

In order to solve~(\ref{sone}) for $\m S\un$,
it is convenient to introduce the following orthonormal basis
of the space of matrices (see~(\ref{psidef}))
\beq
\Psi_{kl,mn}=\psi_{k,m}\psi_{l,n}=\frac{2}{N+1}\sin mq_k\sin nq_l,
\label{Psidef}
\eeq
where the quantised momenta fit to Dirichlet boundary conditions read
\beq
q_k=\frac{k\pi}{N+1}\qquad(k=1,\dots,N).
\label{aqdir}
\eeq
An arbitrary matrix $X_{mn}$ can be expanded as
\beq
X_{mn}=\sum_{kl}\w X_{kl}\Psi_{kl,mn},\qquad
\w X_{kl}=\sum_{mn}X_{mn}\Psi_{kl,mn}.
\eeq
With these notations, we have
\beq
\w B\ze_{kl}=\w B\ze_k\delta_{kl},\qquad
\w S\ze_{kl}=\frac{\delta_{kl}}{\w B\ze_k},
\eeq
where
\beq
\w B\ze_k=1-2K\cos q_k
\eeq
are the eigenvalues of $\m B\ze$ (see~(\ref{eqb})).
Moreover (see~(\ref{belts}))
\beq
\w B\un_{kl}=\frac{2K}{N+1}\sum_m\sin mq_k(\sin(m+1)q_l-\sin(m-1)q_l),
\eeq
where the quantisation of the momenta automatically
takes boundary conditions into account.
The sum over $m$ can be reduced to geometric sums.
We thus obtain
\beq
\w B\un_{kl}=\frac{4K}{N+1}\,\frac{\sin q_k\sin q_l}{\cos q_l-\cos q_k}\qquad(k+l\odd),
\label{b1kl}
\eeq
while these amplitudes vanish if the sum $k+l$ is even.
The solution to~(\ref{sone}) reads
\beq
\w S\un_{kl}=\frac{\w B\ze_l-\w B\ze_k}{\w B\ze_k\w B\ze_l(\w B\ze_k+\w B\ze_l)}\,\w B\un_{kl}.
\label{s1kl}
\eeq
Some algebra using the above results yields the explicit expression
\beq
\Omega=\frac{16K^2}{(N+1)^2}\sum_{k+l\odd}
\frac{\sin^2q_k\sin^2q_l}{(1-K(\cos q_k+\cos q_l))(\cos q_k-\cos q_l)^2}.
\label{aeres1}
\eeq

Some more work is still needed in order to extract the size dependence of this
result in a form similar to~(\ref{bandb}).
In a first step, introducing the odd integers $i=k+l$ and $j=k-l$,
as well as the angles
\beq
\theta_i=\frac{i\pi}{2(N+1)}=\frac{q_k+q_l}{2},\qquad
\theta_j=\frac{j\pi}{2(N+1)}=\frac{q_k-q_l}{2},
\eeq
we can recast~(\ref{aeres1}) as
\beq
\Omega=\frac{K^2}{2(N+1)^2}\sum_{i,j\odd}\frac{1}{1-2K\cos\theta_i\cos\theta_j}
\left(\frac{\sin\theta_i}{\sin\theta_j}-\frac{\sin\theta_j}{\sin\theta_i}\right)^2,
\label{eres2}
\eeq
where each odd integer $i$ or $j$ runs over $2(N+1)$ values between 1 and $4N+3$.
In a second step, expanding~(\ref{eres2}) as a power series in $K$
(i.e., a high-temperature series) yields
\beq
\Omega=\frac{1}{(N+1)^2}\sum_{m\ge0}2^{2m}K^{2m+2}(\alpha_m\beta_{m+1}-\alpha_{m+1}\beta_m),
\label{wab}
\eeq
with
\beq
\alpha_m=\sum_{i\odd}\cos^{2m}\theta_i,\qquad
\beta_m=\sum_{i\odd}\frac{\cos^{2m}\theta_i}{\sin^2\theta_i}.
\eeq
The latter sums enjoy a few simplifying features.
First, the recurrence formula $\beta_{m+1}=\beta_m-\alpha_m$,
with initial value (see~\cite[Eq.(24.1.1)]{hansen})
\beq
\beta_0=\sum_{i\odd}\frac{1}{\sin^2\theta_i}=2(N+1)^2,
\eeq
allows us to express the $\beta_m$ in terms of the $\alpha_m$.
Second, for $N$ large enough,
the sum~$\alpha_m$ is exactly given by the product of the corresponding normalised integral,
\beq
\int_0^{2\pi}\frac{\dd\theta}{2\pi}\cos^{2m}\theta=\frac{(2m)!}{2^{2m}m!^2},
\eeq
and of the number of terms, i.e., $2(N+1)$.
This is a consequence of the Poisson summation formula.
We thus obtain
\beq
\alpha_m=\frac{(2m)!}{2^{2m-1}m!^2}(N+1)\qquad(N\ge m).
\label{apoiss}
\eeq

The above estimate is sufficient to obtain a result of the form~(\ref{bandb})
for the entropy production amplitude on a large open chain, i.e.,
\beq
\Omega\approx N\omega+\Omega\bdy.
\eeq
The entropy production amplitude per spin reads
\beq
\omega=2\sum_{m\ge0}\frac{(2m)!}{m!(m+1)!}K^{2m+2}=1-\sqrt{1-4K^2},
\label{aoser}
\eeq
in agreement with~(\ref{philim}), as $\omega=\varphi/V^2$.
The boundary contribution reads
\beqa
\Omega\bdy&=&\omega-\sum_{m\ge0}\frac{(2m)!(2m+1)!}{2^{2m-2}m!^3(m+1)!}K^{2m+2}
\nonumber\\
&=&\omega+\frac{4}{\pi}({\bf E}(2K)-{\bf K}(2K)),
\label{aomegabdy}
\eeqa
where ${\bf E}$ and ${\bf K}$ are the complete elliptic integrals.
The boundary term $\Omega\bdy$ is always negative.
It diverges logarithmically as
\beq
\Omega\bdy\approx\frac{4}{\pi}\left(\ln\frac{\mu}{4}+1\right)+1
\label{abdylog}
\eeq
as the critical point is approached ($\mu\to0$).

Right at the critical point, the result~(\ref{wab}) simplifies to
\beq
\Omega=N+1-\frac{1}{4(N+1)^2}\sum_{m\ge0}\alpha_m(\alpha_m+\alpha_{m+1}).
\eeq
Moreover, the large-$m$ behavior of the estimate~(\ref{apoiss}), i.e.,
\beq
\alpha_m\approx\frac{2N}{\sqrt{\pi m}},
\eeq
holds up to values of $m$ comparable to $N^2$,
beyond which the sums~$\alpha_m$ fall off exponentially fast,
as $\exp(-\pi^2m/(4N^2))$.
We thus obtain
\beq
\Omega\approx N-\frac{4}{\pi}\ln\frac{N}{N_1}.
\label{acritlog}
\eeq
The finite part $N_1$ is not predicted by the above line of reasoning.

\section{Derivation of the expressions~(\ref{omegares}) and~(\ref{rcomegares})
for the amplitude matrices of the $RL$ and $RC$ arrays}
\label{aelec}

In this appendix we give a full derivation of the expressions~(\ref{omegares})
and~(\ref{rcomegares}) for the amplitude matrices $\m\Omega$
of the $RL$ and $RC$ electrical networks investigated in section~\ref{elec},
for an arbitrary network size $N$.

\subsection{$RL$ networks (see section~\ref{rl})}
\label{arl}

The derivation of the amplitude matrix $\m\Omega$
entering~(\ref{ophidef}) goes as follows.
The expression~(\ref{telts}) of the temperature matrix can be recast as
\beq
\m\T=\Tbar\m\Delta+\m\eps,
\eeq
where the non-zero elements of the matrix $\m\eps$ are
\beqa
\eps_{nn}=\delta T_{n-1}+\delta T_n,
\nonumber\\
\eps_{n,n-1}=\eps_{n-1,n}=-\delta T_{n-1}\qquad(n\ne1).
\label{epselts}
\eeqa
Along the lines of section~\ref{smallv},
let us evaluate the matrices $\m S$ and $\m Q$ to first order in~$\m\eps$,
i.e., to first order in the temperature fluctuations $\delta T_n$.
Looking for a solution to the Sylvester equation~(\ref{sidrl}) of the form
\beq
\m S\approx\frac{1}{L}(\Tbar\,\m 1+\m\sigma),
\eeq
we find that the matrix $\m\sigma$ obeys
\beq
\m\Delta\m\sigma+\m\sigma\m\Delta=2\m\eps.
\label{sigmaeq}
\eeq
We have then
\beq
\m Q\approx\frac{R}{L^2}(\m\Delta\m\sigma-\m\eps)
\approx\frac{R}{L^2}(\m\eps-\m\sigma\m\Delta).
\label{qrl}
\eeq
Finally, using~(\ref{phi2}), we obtain the following expression
\beq
\Phi\approx\frac{R}{L\Tbar^2}\tr(\m\Delta^{-1}\m\eps(\m\Delta\m\sigma-\m\eps))
\label{phirl}
\eeq
for the entropy production rate to the required accuracy,
i.e., to second order in the temperature fluctuations $\delta T_n$.

In order to solve~(\ref{sigmaeq}) for $\m\sigma$,
it is convenient to use the same formalism as in~\ref{asmallv},
including the quantised momenta~(\ref{aqdir})
and the orthonormal basis introduced in~(\ref{Psidef}).
Within these conventions, we have
\beq
\w\Delta_{kl}=2(1-\cos q_k)\delta_{kl},
\eeq
whereas~(\ref{epselts}) yields after some algebra
\beq
\w\eps_{kl}=\frac{8}{N+1}\sin\half q_k\sin\half q_l\sum_{n=0}^N
\cos(n+\half)q_k\cos(n+\half)q_l\,\delta T_n.
\label{epskl}
\eeq
The solution to~(\ref{sigmaeq}) then reads
\beq
\w\sigma_{kl}=\frac{\w\eps_{kl}}{2-\cos q_k-\cos q_l}.
\label{sigkl}
\eeq
As a consequence, (\ref{phirl}) can be recast as
\beq
\Phi\approx\frac{R}{4L\Tbar^2}
\sum_{kl}\frac{(\cos q_k-\cos q_l)^2(\w\eps_{kl})^2}{(1-\cos q_k)(1-\cos q_l)(2-\cos q_k-\cos q_l)}.
\eeq
Finally, using~(\ref{epskl}),
this expression can be brought to the announced form~(\ref{ophidef}),
where the entries of the amplitude matrix $\m\Omega$ read
\beqa
\Omega_{mn}&=&\frac{4}{(N+1)^2}
\sum_{kl}\frac{(\cos q_k-\cos q_l)^2}{2-\cos q_k-\cos q_l}
\nonumber\\
&\times&\cos(m+\half)q_k\cos(m+\half)q_l\cos(n+\half)q_k\cos(n+\half)q_l.
\label{aomegares}
\eeqa

\subsection{$RC$ networks (see section~\ref{rc})}
\label{arc}

The derivation of the amplitude matrix $\m\Omega$
entering~(\ref{rcodef}) follows the same lines.
The duality symmetry~(\ref{sdual}) implies that the covariance matrix is given by
\beq
\m S\approx C(\Tbar\,\m 1+\m\sigma),
\eeq
where the matrices $\m\eps$ and $\m\sigma$ are the same as in~\ref{arl},
whereas~(\ref{qrl}) and~(\ref{phirl}) respectively become
\beq
\m Q\approx\frac{1}{R}\m\Delta^{-1}(\m\sigma\m\Delta-\m\eps)\m\Delta^{-1}
\approx\frac{1}{R}\m\Delta^{-1}(\m\eps-\m\Delta\m\sigma)\m\Delta^{-1}
\eeq
and
\beq
\Phi\approx\frac{1}{RC\Tbar^2}
\tr(\m\Delta^{-1}\m\eps\m\Delta^{-2}(\m\sigma\m\Delta-\m\eps)).
\label{phirc}
\eeq
Using~(\ref{sigkl}), the latter expression can be recast as
\beq
\Phi\approx\frac{1}{16RC\Tbar^2}
\sum_{kl}\frac{(\cos q_k-\cos q_l)^2(\w\eps_{kl})^2}{(1-\cos q_k)^2(1-\cos q_l)^2(2-\cos q_k-\cos q_l)}\!.
\eeq
Finally, using~(\ref{epskl}),
this expression can be brought to the announced form~(\ref{rcodef}),
where the entries of the amplitude matrix $\m\Omega$ read
\beqa
\Omega_{mn}&=&\frac{1}{(N+1)^2}
\sum_{kl}\frac{(\cos q_k-\cos q_l)^2}{(1-\cos q_k)(1-\cos q_l)(2-\cos q_k-\cos q_l)}
\nonumber\\
&\times&\cos(m+\half)q_k\cos(m+\half)q_l\cos(n+\half)q_k\cos(n+\half)q_l.
\label{arcomegares}
\eeqa

\end{document}